\documentclass[final,5p,authoryear]{elsarticle}

\usepackage{amssymb}
\usepackage{amsmath}
\usepackage{subfig}
\usepackage{enumitem}
\usepackage{xcolor}
\usepackage{lineno}
\usepackage{hyperref}
\usepackage{booktabs}
\usepackage{multirow}
\usepackage{physics}
\journal{Earth and Planetary Science Letters}

\begin{document}

\begin{frontmatter}

 \title{Hydrodynamical simulations of proto-Moon degassing}
 \author[a]{G. Madeira}
 \ead{madeira@ipgp.fr}
 \affiliation[a]{organization={Université Paris Cité, Institut de Physique du Globe de Paris, CNRS},
            city={Paris},
            postcode={F-75005}, 
            country={France}}
 \affiliation[b]{organization={Université Côte d’Azur, Observatoire de la Côte d’Azur, CNRS, Lagrange (UMR7293)},
            city={Nice cedex 4},
            postcode={F-06304}, 
            country={France}}
 \affiliation[c]{organization={São Paulo State University, Grupo de Dinâmica Orbital e Planetologia},
            city={Guaratinguetá, SP},
            postcode={12516-410}, 
            country={Brazil}}
\author[a,c]{L. Esteves}
\author[a]{S. Charnoz}
\author[b]{E. Lega}
\author[a]{F. Moynier}


\begin{abstract}
Modeling the isotopic and elemental abundance of the bulk silicate Moon represent major challenges. Similarities in the non-mass dependent isotopic composition of refractory elements with the bulk silicate Earth suggest that both the Earth and the Moon formed from the same material reservoir. On the other hand, the Moon's volatile depletion and isotopic composition of moderately volatile elements points to a global devolatilization processes, most likely during a magma ocean phase of the Moon.  Here, we investigate the devolatilisation of the molten Moon due to a tidally-assisted hydrodynamic escape, first proposed by \cite{charnoz2021tidal}, with a focus on the dynamics of the evaporated gas. Unlike the 1D steady-state approach of \cite{charnoz2021tidal}, we use 2D time-dependent hydrodynamic simulations carried out with the \texttt{FARGOCA} code modified to take into account the magma ocean as a gas source. Near the Earth's Roche limit, where the proto-Moon likely formed, evaporated gases from the lunar magma ocean form a circum-Earth disk of volatiles, with less than 30\% of material being re-accreted by the Moon. We find that the measured depletion of K and Na on the Moon can be achieved if the lunar magma-ocean had a surface temperature of about 1800-2000 K. After about 1000 years, a thermal boundary layer or a flotation crust forms a lid that inhibits volatile escape. Mapping the volatile velocity field reveals varying trends in the longitudes of volatile reaccretion on the Moon's surface: material is predominantly re-accreted on the trailing side when the Moon-Earth distance exceeds 3.5 Earth radii. For $k_2/Q$ values of 0.0003 and 0.03, 60\% and more than 99\% of the volatile material, respectively, is re-accreted on the trailing side, suggesting a dichotomy in volatile abundances between the leading and trailing sides of the Moon. This dichotomy may provide insights on the tidal conditions of the early molten Earth. In conclusion, tidally-driven atmospheric escape effectively devolatilizes the Moon, matching the measured abundances of Na and K on timescales compatible with the formation of a thermal boundary layer or an anorthite flotation crust.
\end{abstract}



\end{frontmatter}

\section{Introduction} \label{sec:intro}

The similarities in the isotopic composition of oxygen \citep{wiechert2001oxygen,herwartz2014identification} and non-volatile elements such as Cr and Ti \citep{zhang2012proto,mougel2018chromium} between the bulk silicate Earth and the Moon indicate that both Earth's mantle and the bulk silicate Moon originated from material with a similar initial isotopic composition. The Moon's formation scenario is still debated: The 'canonical scenario', involving an impact with a Mars-sized object \citep{benz1989origin, kokubo2000lunar, canup2001origin, canup2008lunar}, successfully explains the anomalously high angular momentum of the Earth-Moon system and the Moon's bulk iron depletion. However, it reproduces the compositional similarities between the Earth and the Moon only if the Earth and the impactor shared nearly identical compositions. This limitation arises because the simulations of the canonical scenario obtain a proto-lunar disk predominantly composed of impact material. Nevertheless, the formation of planet-sized objects with the required compositional similarities is highly improbable \citep{kaib2015brief, kaib2015feeding}.

High-energy, high-angular-momentum impacts \citep{cuk2012making,canup2012forming,lock2018origin,lock2020energy} produce a higher-temperature proto-lunar disk, where the homogenization of the proto-Earth and the impactor material is facilitated. However, the substantial vapor mass above the proto-lunar disk produced by these impacts can, in fact, inhibit the formation of a satellite as massive as the Moon \citep{nakajima2024limited}, which has been shown to be a caveat of these high-energy models. Additionally, achieving the level of material mixing necessary to explain the Earth-Moon system requires a significant excess of angular momentum \citep{meier2024systematic}.

\cite{cuk2012making} propose that an evection resonance with the Sun might have removed angular momentum from the post-impact system; however, it remains uncertain whether this mechanism alone can sufficiently account for the removal of all the angular momentum excess \citep{rufu2020tidal, ward2020analytical}. Regardless of the scenario, the physics governing the debris disk, the recondensation and reaccretion of evaporated silicates into condensates, and their coupled dynamics remain poorly understood due to the intrinsic complexity of the problem, leaving the Moon's formation as an open question.

In contrast to the similar abundance of refractory elements between the Earth and the Moon, the abundances of moderately volatile elements (e.g., Na, K, Zn, Rb, Cd) in lunar mare basalts exhibit a notable depletion when compared to terrestrial basalts \citep{ringwood1992volatile,lodders2015chemistry,day2014evaporative}. Furthermore, it has been measured that volatile elements, including Zn \citep{paniello2012zinc,kato2015extensive,day2022partial}, K \citep{wang2016potassium,tian2020potassium}, Cl \citep{sharp2007chlorine,boyce2015chlorine,gargano2020cl}, Ga \citep{kato2017gallium}, and Rb \citep{pringle2017rubidium}, are enriched in the heavier isotopes in the Moon compared to the Earth, suggesting that the depletion of volatile elements in the Moon is attributable to global evaporation processes.

High-precision analysis of lunar magmatic rocks \citep{sossi2018volatile} reveals an enrichment of lighter Cr isotopes compared to Earth's mantle, suggesting that Cr partitioning occurred near equilibrium in an oxygen-rich environment, at relatively low temperatures of around 1600–1800~K. \cite{sossi2018volatile} argue that these temperatures are similar to those expected from degassing at the surface of a magma ocean, indicating that volatile loss likely occurred during the magma ocean phase. However, we point out that it is also possible this process took place right after the giant impact, after the gas expanded and cooled to low temperatures. Nevertheless, due to the absence of hydrogen and the Earth's large mass, it seems unlikely that moderately volatile elements were lost through thermal escape \citep{Nakajima_Stevenson_2018}. This is why several alternative scenarios have been proposed.

In the protolunar disk, volatile loss could happen due to incomplete condensation and liquid-vapor partitioning, caused by the viscous separation of the vapor from the liquid \citep{charnoz2015evolution} or by the gravitational torque exerted on the vapor by the proto-Moon \citep{canup2015lunar}. Additionally, \cite{lock2018origin,lock2020energy} propose that separation due to the gas drag of the growing droplets from the vapor may also play a role. However, the effectiveness of these scenarios depends heavily on the structure of the protolunar disk, which remains highly unknown. In fact, the existence of a  protolunar disk is even debated. For example, \cite{kegerreis2022immediate} demonstrates through high-resolution simulations that the Moon may have formed immediately from a high-energy oblique impact of an object with the proto-Earth \citep[also see][]{meier2024systematic}.

A plausible explanation is that the volatile loss occurred later in the system, after the Moon underwent wholesale melting following a high-energy accretion process and became covered by an ocean of magma \citep{wood1970lunar}. Isotopic evidence supporting this scenario comes from the variable Zn and Cl isotopic composition between lunar lithologies \citep{kato2015extensive,boyce2015chlorine}. However, while isotopic data points to volatile loss during magma ocean degassing, the Moon's escape velocity is too large and would prevent the loss of moderately volatile elements.

To solve this conundrum \cite{charnoz2021tidal} developed a new model in which the volatile loss occurred immediately after the assembly of the proto-Moon, when the object was much closer to Earth. The influence of Earth's tides reduces the velocity needed for a gas to cross the Hill sphere and thermally escape from the proto-Moon. Exploring this scenario, \cite{charnoz2021tidal} find that between 3 and 6 Earth radii, the proto-Moon undergoes intense tidal devolatilization, and volatile material does not re-enter the satellite’s Hill sphere when under perturbative effects such as gas drag or radiation pressure. For lunar surface temperatures of approximately 1600-1800 K, \cite{charnoz2021tidal} demonstrated that lunar depletion in Na and K can be reproduced, provided the escape persists for at least 1000 years.

However, \cite{charnoz2021tidal} does not fully evaluate the problem by assuming a 1D steady-state approach. They solve the classical equations of hydrodynamic escape \citep{parker1963kinematical,parker1965dynamical} only for the line connecting the Earth and the Moon, assessing whether the gas on the surface of the proto-Moon is hot enough to accelerate upwards to the Lagrange points, which are assumed to be situated on the Earth-Moon line. Consequently, their calculations of fluxes account for material released from only a limited region of the lunar surface. Moreover, they fail to systematically demonstrate that the material does not return to the Moon, relying instead on simulations with test particles that do not consider the hydrodynamic nature of the gas evolution. In fact, as we will demonstrate below, the presence of gas in the vicinity of the Moon acts to displace the Lagrange points away from the Earth-Moon line, challenging an important assumption of their model.

In this work, we take a step further than \cite{charnoz2021tidal} and investigate the tidally-driven atmospheric escape scenario using 2D time-dependent hydrodynamic simulations. This approach allows us to evaluate the evolution of gas released from the Moon's equator, quantifying both the amount of material that returns to the satellite and the locations on its surface where the gas is re-accreted. Additionally, we quantify the flow of material from the Moon to the Earth. In Section~\ref{sec:methods}, we present our hydrodynamic set-up and our model for the Moon's magma ocean. In Section~\ref{sec:circumearth}, we numerically simulate the evolution of gas released from the Moon and calculate the material fluxes. In Section~\ref{sec:finalcomposition}, we examine the Moon's orbital evolution due to tides, calculating how the abundances of Na and K in the satellite change over time. Then, in Section~\ref{sec:lid}, we explore how the formation of a stagnant lid can influence the total amount of material lost by the Moon. We discuss our results in Section~\ref{sec:discussion} and draw our main conclusions in Section~\ref{sec:conclusion}.

\section{Methods} \label{sec:methods}
Our 2D numerical simulations were conducted using the \texttt{FARGOCA} code \citep[FARGO with Co-latitude Added,]{lega2014migration}\footnote{The simulations presented in this paper were obtained with a recently refactored version of the code available at \url{https://gitlab.oca.eu/DISC/fargOCA}}. \texttt{FARGOCA} is a parallelized, grid-based hydrodynamic code originally designed for studying protoplanetary disks. It is based on the \texttt{FARGO} code \citep{masset2000fargo} and solves the fluid equations using a second-order upwind scheme within a time-explicit-implicit multi-step procedure.

We use polar coordinates $(r,\theta)$, where $r$ is the radial distance from the Earth (which is at the origin of the coordinate system) and $\theta$ is the azimuthal coordinate measured from the $x$-axis. We work in a coordinate system that rotates with the angular velocity of the Moon:
\begin{equation}
\Omega_M = \sqrt {\frac{G(M_E+M_M)}{{a_M}^3}}
\end{equation}
where $M_E$ is the mass of the Earth, $M_M$ is the current mass of the Moon, $G$ is the gravitational constant, and $a_M$ is the semi-major axis of the Moon. The Moon is assumed to be in a circular and planar spin-tidally locked orbit.

The dynamics of the gas are described by the integration of the Navier-Stokes equations, composed of a continuity equation and a set of two momentum equations:

\begin{itemize}[label={},leftmargin=0pt]
\item {\bf Continuity equation}\\
\hspace*{1em} The continuity equation is given by:
\begin{equation}
\label{continuity}
\frac{\partial \Sigma} { \partial t}+ \nabla \cdot (\Sigma \vec v)= 0
\end{equation}
where $\Sigma$ is the surface density of the gas and $\vec v=(v_r,v_\theta)$
is the velocity vector of the gas flow, with $v_\theta=r(\omega+\Omega_M)$ where $\omega$ is the azimuthal angular velocity in the rotating frame.

\item {\bf Equations for the momenta}\\
\hspace*{1em} The Navier-Stokes equations for the radial momentum $J_r= \Sigma v_r$, and the angular momentum $J_\theta = \Sigma r v_\theta $ read:
\begin{equation}
\label{navstocons3D}
\left\lbrace \begin{array}{lll}
{\partial J_r \over \partial t}+ \nabla \cdot (J_r\vec v) & = & \Sigma \left[\frac{v_\theta^2}{r} - \frac{\partial \Phi}{\partial r} + \frac{1}{\Sigma} \left(f_r - \frac{\partial P_{\rm 2D}}{\partial r}\right)\right] \\
{\partial J_\theta \over \partial t}+ \nabla \cdot (J_\theta \vec v) & = &
\Sigma r \left[-\frac{\partial \Phi}{\partial \theta} + \frac{1}{\Sigma} \left(f_\theta - \frac{\partial P_{\rm 2D}}{\partial \theta}\right)\right]
\end{array} \right.
\end{equation}

where $P_{\rm 2D}$ is the vertically integrated pressure. The function $f=(f_r,f_\theta)$, which is the divergence of the stress tensor, is proportional to the disk's viscosity. In this study, we consider the fluid inviscid based on the results of previous one-dimensional calculations \citep{charnoz2021tidal}. Although an inviscid fluid is an appropriate approximation for low-density gases, as is the case for our vapor gas (Section~\ref{sub_composition}), we point out that at the magma ocean interface, the gas density, and consequently its viscosity, may reach higher values, representing a caveat of our model.

The potential $\Phi$ acting on the disk consists of the contribution of the Earth $\Phi_E=-GM_{E}/r$ and the Moon $\Phi_M$, plus indirect terms that arise from the primary acceleration due to the Moon's and disk’s gravity \citep{Masset2002}. The gravitational potential of the Moon on the disk is modeled as:
\begin{equation}
\Phi_M = -\frac{GM_M}{\sqrt{r^2+R_{sm}^2}}
\end{equation}
where $R_{sm}$ is the smoothing length with $R_{sm}=0.6H$, and $H=h r$ is the disk scale height, $h$ being the disk's aspect ratio (whose value is given in Section 2.1).

\item {\bf Energy equation}\\
\hspace*{1em} To the Navier-Stokes system of equations, we add the following equation for the temporal variation of the 2D internal energy density:
\begin{equation}
\frac{\partial e}{\partial t} + \nabla \cdot (e\vec{v}) = -P_{\rm 2D} \nabla \cdot \vec{v},
\end{equation}
where the term \(-P_{\rm 2D} \nabla \cdot \vec{v}\) is the compressional heating, and the internal energy density is given by
\begin{equation}
e = \Sigma c_v T, \label{eq_e}
\end{equation}
with \(T\) being the gas temperature and \(c_v\) the specific heat at constant volume.

The vertically integrated pressure relates to the 2D internal energy density through the equation
\begin{equation}
P_{\rm 2D} = (\gamma - 1)e, \label{eq_P}
\end{equation}
where \(\gamma\) is the adiabatic index.
\end{itemize}

In this work, our aim is to explore the dynamics of gas released from the surface of the Moon, deviating slightly from the scenarios for which the code was originally intended. Specifically, the \texttt{FARGOCA} code employs a point-mass approach for calculating the gravitational force of planets -- or satellites, in our case -- which is not appropriate for our simulations that require a physically resolved surface for the Moon. For this reason, we implemented modifications in the code, described below.

\subsection{Implementations} \label{sec:impl}
We implemented the extended dimensions of the Moon in the code using boundary conditions. For each integration time-step, we calculate the distance \(d_{i,M}\) from each grid cell \(i\) to the Moon's center of mass. Cells within a given sphere contained by the Moon are assumed to have the same properties as the Moon's center. Meanwhile, an evanescent boundary condition \citep{de2006comparative} is applied in the narrow shell between this sphere and the Moon's surface. This boundary condition is mathematically defined as:
\begin{equation}
\Xi = \begin{cases} 
\Xi_M & \text{for~} d_{i,M} \leq R_{\mathrm{shell}}, \\
\frac{\Xi + \lambda \Xi_M}{1 + \lambda} & \text{for~} R_{\mathrm{shell}} \leq d_{i,M} \leq R_M 
\end{cases} \label{eq_bc}
\end{equation}
where \(\Xi\) represents any of the following quantities: velocity components, surface density, and energy. \(\Xi_M\) is the quantity at the Moon's center, and \(\lambda\), as defined by \cite{de2006comparative}, is the evanescent parameter responsible for controlling the rate of damping. The parameter relates to the integration time-step \(T_{\rm step}\) through the equation:
\begin{equation}
\lambda = \frac{20 T_{\rm step}}{\mathcal{T}_M}\left(\frac{R_M - d_{i,M}}{R_M - R_{\mathrm{shell}}}\right)^2,
\end{equation}
where \(\mathcal{T}_M\) is the Moon's orbital period.

The evanescent boundary condition applied to the Moon's surface is based on the approach originally used for the edges of the disk in the \texttt{FARGOCA} code. It is explicitly applied to the velocity, energy, and surface density of the gas. At the Moon's center, the radial and azimuthal components of the velocity in the rotating frame are set to \(v_{r,M} = v_{\theta,M} = 0\), and the surface density is computed as (see Equations~\ref{eq_e}-\ref{eq_P}):
\begin{equation}
\Sigma_M = \frac{P_{\rm 2D}}{T_M} \frac{1}{c_v(\gamma - 1)},  \label{eq:sigma}
\end{equation}
where \(T_M\) is the temperature at the Moon's center. The gas energy at the Moon's center is computed using Equation~\ref{eq_e}. It should be noted that the condition \(T = T_M\) at the Moon's center is applied implicitly in our code.

This approach ensures that all gas material within the sphere behaves as a single entity while promoting a smooth transition between the regions inside and outside the sphere. Simultaneously, if the surface density at the sphere, $\Sigma_M$, is higher than the disk's local surface density $\Sigma$, we naturally observe a flow of material from the Moon's surface to the disk. We interpret this as the vapor flux produced in the magma ocean.

\subsection{Numerical setup-up}
Our numerical simulations include the Earth as the central object, and the Moon, modeled following the implementation described in Section~\ref{sec:impl}. Our calculations are performed in normalized units, assuming the Earth mass $M_E$ and radius $R_E$ as units of mass and distance, respectively (G=$M_E$=$R_E$=1). We model the environment around the Earth using 1000 grid cells with arithmetic spacing in radius ($\delta r=0.011$), distributed from 1 to 12 $R_E$, and 1000 grid cells in azimuth for the full circumference ($\delta \phi=0.001$). This configuration results in a total of 10224 cells representing the Moon. 

To ensure accuracy, we ran test simulations spanning 0.1~yr, measuring the fluxes of material ejected, re-accreted by the Moon, and lost at the disk boundaries. The outer disk boundary was increased from 9 $R_E$ in steps of 1 $R_E$, and grid resolution was refined from 500 to 1000 cells in both radial and azimuthal directions, in steps of 100. Fluxes converged to asymptotic values with increasing precision. We adopted a disk edge of 12 $R_E$ and 1000 grid cells, as these configurations resulted in relative flux differences below 10\% compared to the immediately less precise cases. We also performed a set of test simulations varying the values of $R_{\mathrm{shell}}$ to check the convergence of results with respect to this parameter, in the case of the Moon positioned at 3~$R_E$. These tests revealed that the results consistently converged for a minimum value of 0.95~$R_M$, which we adopted as the standard $R_{\mathrm{shell}}$ in our study. The disk's inner and outer boundaries were treated as open.

Ideally, our simulations should begin with only the Earth and Moon present, to accurately analyze the flow of volatiles in the system. However, due to a code limitation, grid cells cannot be initialized as empty. For simplicity, we start our simulations with an inviscid, adiabatic disk that has a constant surface density of $\Sigma=1.6\times10^{-15}$~g/cm$^2$. This density is close to the minimum the code can handle and corresponds to a total mass of less than $<10^{-22}$ $M_E$. Moreover, this value is at least four orders of magnitude smaller than the $\Sigma_M$ values assumed in our simulations. The disk is characterized by a constant aspect ratio $h$, computed in normalized units as:
\begin{equation}
h=\sqrt{\frac{3RT_{BG}}{\mu}}    
\end{equation}
where $R$ is the gas universal constant, $\mu$ is the gas mean molecular mass, and $T_{BG}$ is the disk background temperature, assumed to be 300~K at 3 $R_E$. Next, we detail our assumptions for the gas vapor released from the Moon's surface.

\subsection{Magma ocean composition} \label{sub_composition}
We assume that the vapor above the lunar magma ocean consists of gas evaporated congruently by the oxides present in a liquid with a bulk silicate Earth (BSE) composition. The vapor composition is computed using the recently published \texttt{VAPOROCK} code \citep{Wolf2023}, that is an evolution of the \texttt{MELT} code. As in \cite{Visscher_Fegley_2013, charnoz2021tidal, Lavatmos_2023}, the fraction of ${\rm O_2}$ in the vapor is computed by assuming mass conservation of oxygen between the magma ocean and the vapor during the vaporization process.

\begin{figure}
    \centering
    \includegraphics[width=1.0\columnwidth]{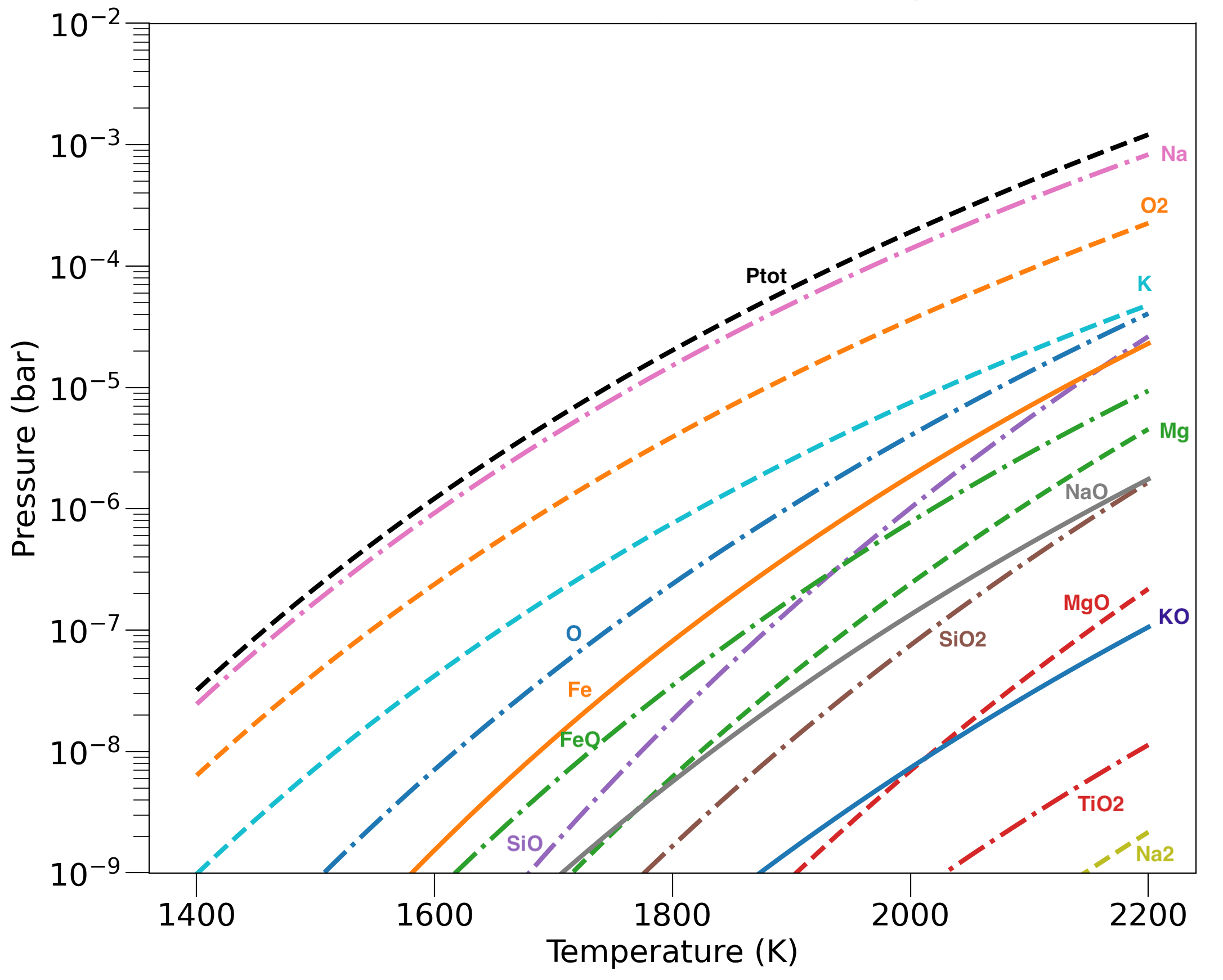}
    \caption{Partial pressure in the vapor above the lunar magma ocean as a function of lunar surface temperature, assuming a bulk silicate Earth (BSE) composition. Calculation was performed using the \texttt{VAPOROCK} code. $Ptot$ stands for the total saturating vapor pressure.}
    \label{fig:vapor_compo}
\end{figure}

Figure \ref{fig:vapor_compo} displays the resulting vapor composition in terms of partial pressures. Our bulk silicate Earth (BSE) vapor composition is very similar to that obtained with the \texttt{LAVATMOS} code \citep[see Figure 4 of][]{Lavatmos_2023} and the one published in \cite{Ito_2015}. However, significant differences appear when compared to the composition reported in \cite{charnoz2021tidal}: our Na and K partial pressures computed with \texttt{VAPOROCK} are respectively about 5 and 10 times smaller at 2000~K than those in \cite{charnoz2021tidal}, due to different models for activity coefficients. Note that, whereas our \texttt{VAPOROCK} calculation is in good agreement with \texttt{LAVATMOS} and \cite{Ito_2015}, there are still notable differences with the vapor composition obtained with the \texttt{MAGMA} code \citep[see Figure 4 of][]{Lavatmos_2023}, which reports about an order of magnitude less Na and two orders of magnitude less K abundances at 2000~K than our calculations, probably due to different thermo-chemical data.

\begin{table*}[]
\centering
\caption{Total pressure, adiabatic index, mean molecular mass, and atomic mass fraction of the lunar magma ocean, initially assumed to have the same composition as the bulk silicate Earth. \label{tab:bse}}
\begin{tabular}{lcccccc}
\hline\hline
   & Symbol     & 1400 K    & 1600 K   & 1800 K   & 2000 K   & 2200 K              \\
Total pressure (bar)        & P          & 3.2$\times 10^{-8}$ & 1.2$\times 10^{-6}$ & 2.0$\times 10^{-5}$ & 1.9$\times 10^{-4}$ & 1.2$\times 10^{-3}$ \\
adiabatic index             & $\gamma$   &      1.3552        & 1.2806              & 1.2278              & 1.1938              & 1.1725              \\
mean molecular mass (g/mol) & $\mu$      &      25.3         & 25.4                & 25.5                & 25.9                & 26.6                \\ \hline
atomic mass fraction        &            &                     &                     &                     &                     &                     \\
O                           & $\mu_{\rm O}$  & 0.253               & 0.252               & 0.252               & 0.254               & 0.264               \\
Si                          & $\mu_{\rm Si}$ & \textless{}0.001    & \textless{}0.001    & 0.001               & 0.006               & 0.024               \\
Mg                          & $\mu_{\rm Mg}$ & \textless{}0.001    & \textless{}0.001    & \textless{}0.001    & 0.001               & 0.004               \\
Fe                          & $\mu_{\rm Fe}$ & 0.001               & 0.004               & 0.013               & 0.030               & 0.056               \\
Na                          & $\mu_{\rm Na}$ & 0.698               & 0.689               & 0.675               & 0.648               & 0.593               \\
K                           & $\mu_{\rm K}$  & 0.048               & 0.054               & 0.059               & 0.061               & 0.059              \\ \hline 
\end{tabular}
\end{table*}

Table~\ref{tab:bse} shows the total pressure, the adiabatic index, the mean molecular mass, and atomic mass fractions of the gas above the magma ocean as functions of the lunar surface temperature. Aluminum (Al) and Calcium (Ca) are also included in our gas composition; however, their mass fraction in the vapor are always less than 0.001 for all temperatures considered. In this work, we conducted a total of 65 numerical simulations, each with a timespan of 1~year, considering the magma ocean temperature \(T_M\) and the Moon's orbital distance \(a_M\) as free parameters. We assume $T_M$=1400, 1600, 1800, 2000, and 2200~K and $a_M$ ranging from 3 to 9 $R_E$, with a semi-major axis step of 0.5 $R_E$. As the Moon loses volatiles, the atomic concentrations in the magma ocean are expected to change over time, affecting the pressure, adiabatic index, and mean molecular mass of the vapor above the magma. For simplicity, however, we treat these parameters as constants throughout the simulations, determined solely by the gas temperature (Table~\ref{tab:bse}), which is also kept constant.

\section{Circum-Earth volatile disk and its dependency on lunar position and surface temperature} \label{sec:circumearth}

\begin{figure*}[]
\centering
\subfloat[]{\includegraphics[width=0.6\columnwidth]{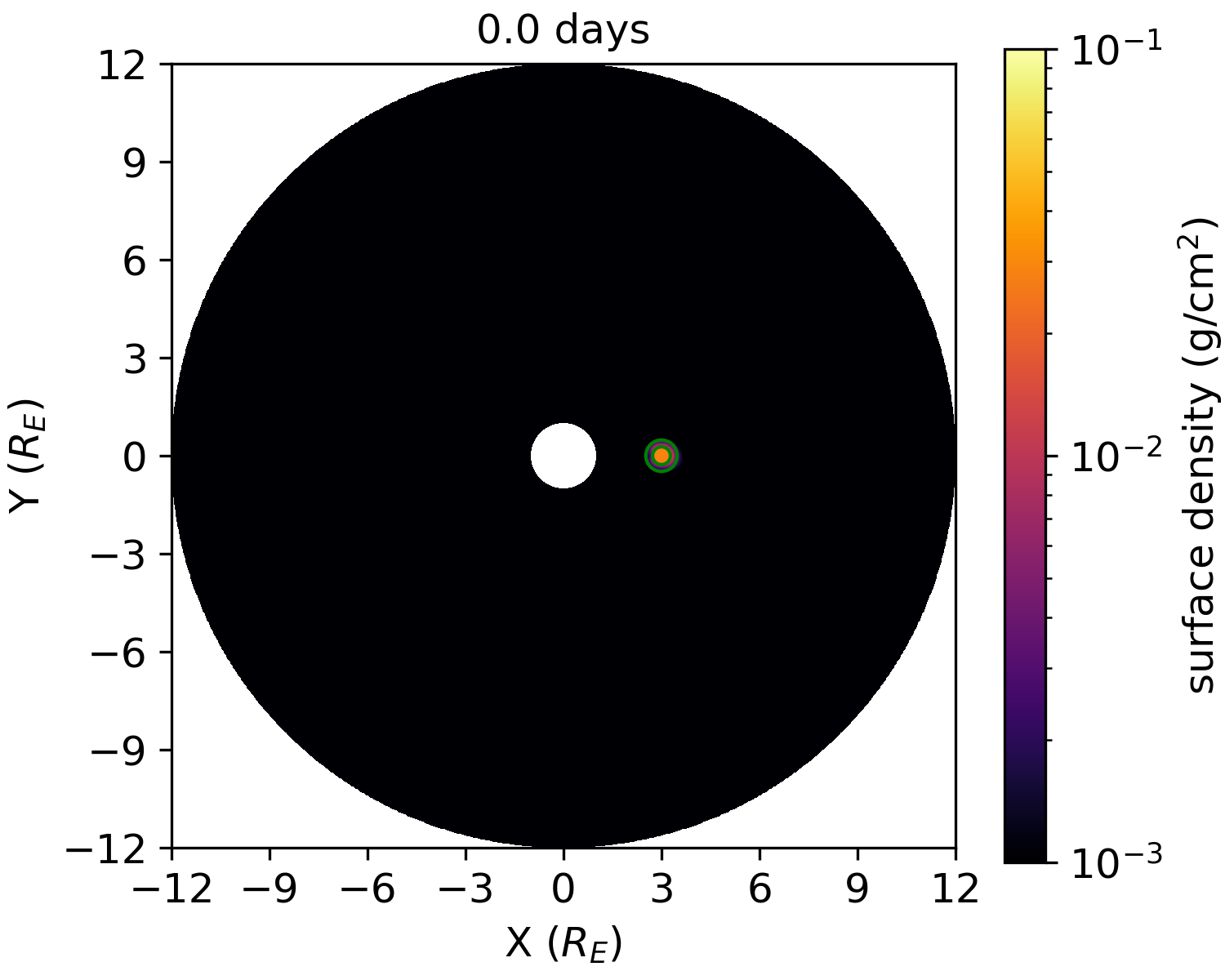}\label{fig:xyminora} }
\quad
\subfloat[]{\includegraphics[width=0.6\columnwidth]{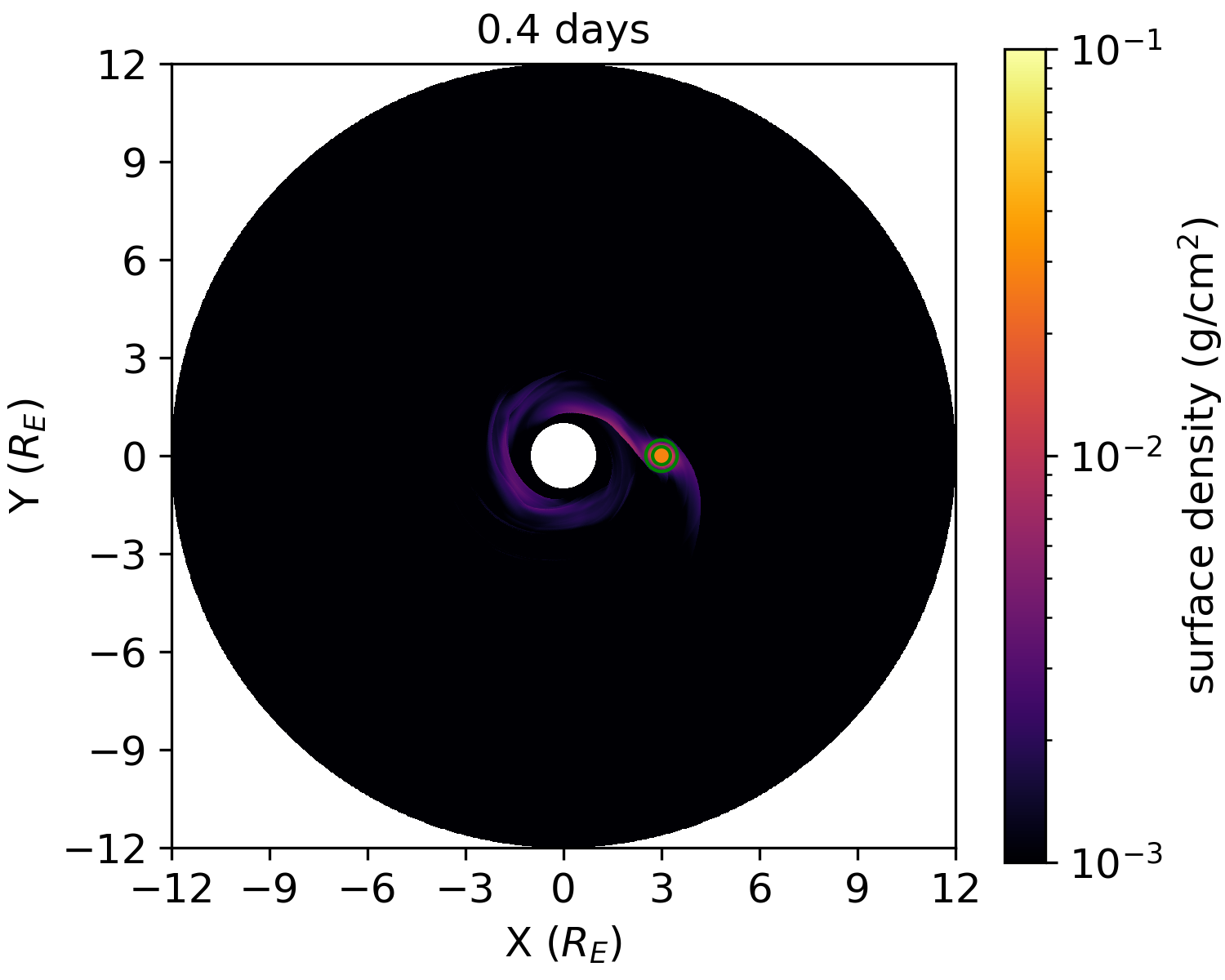}\label{fig:xyminorb}}
\quad
\subfloat[]{\includegraphics[width=0.6\columnwidth]{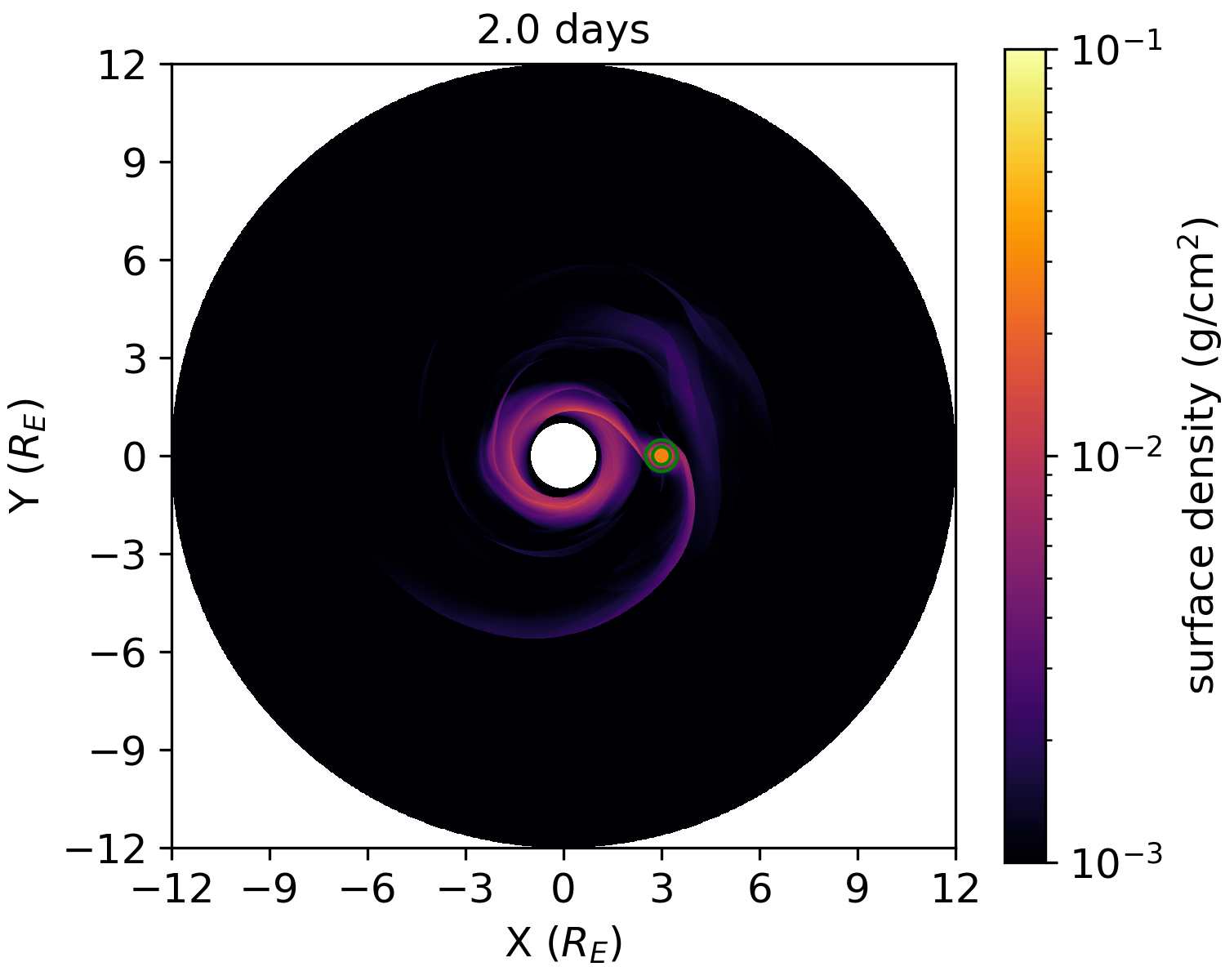}\label{fig:xyminorc}} \\
\centering
\subfloat[]{\includegraphics[width=0.6\columnwidth]{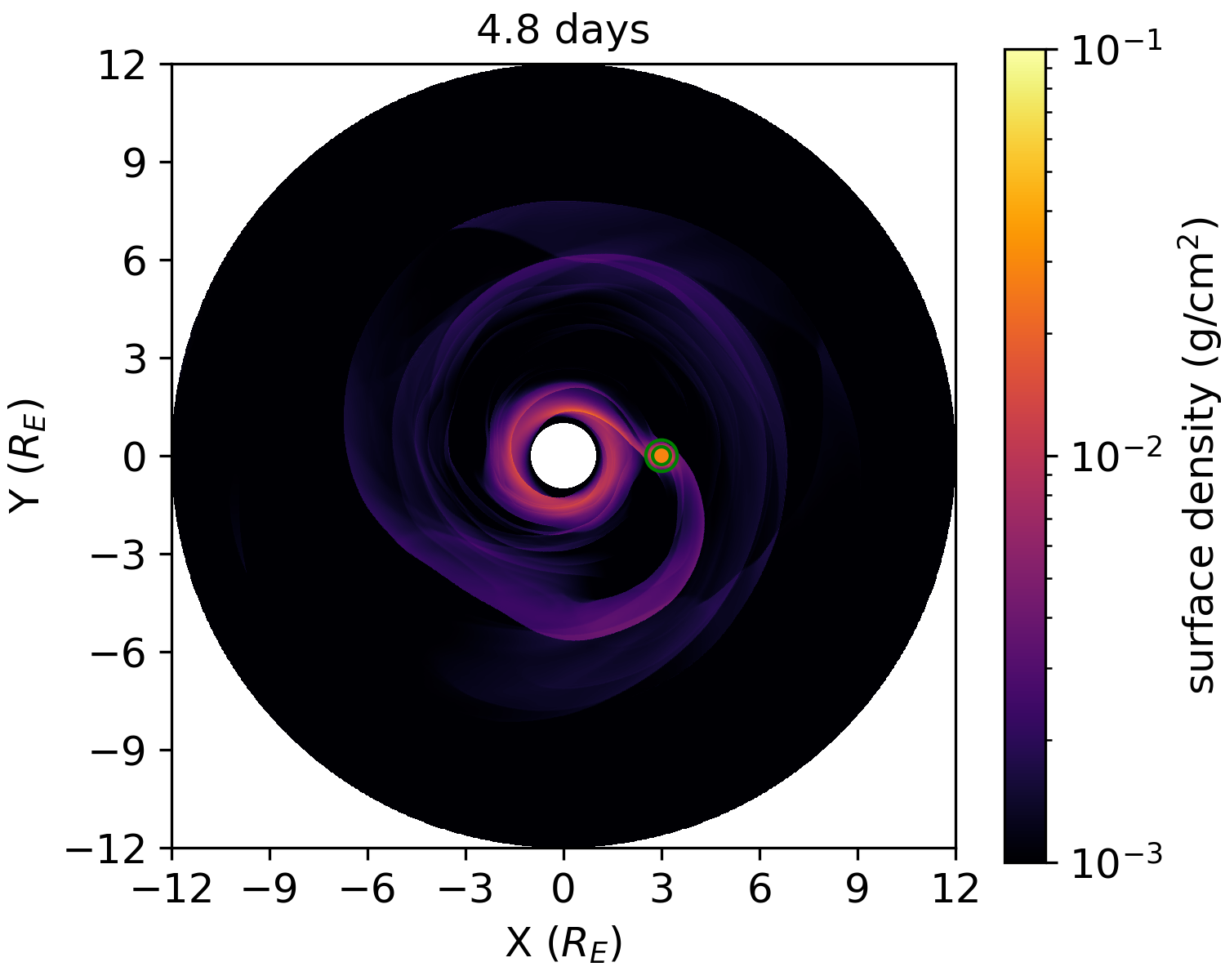}\label{fig:xyminord}}
\quad
\subfloat[]{\includegraphics[width=0.6\columnwidth]{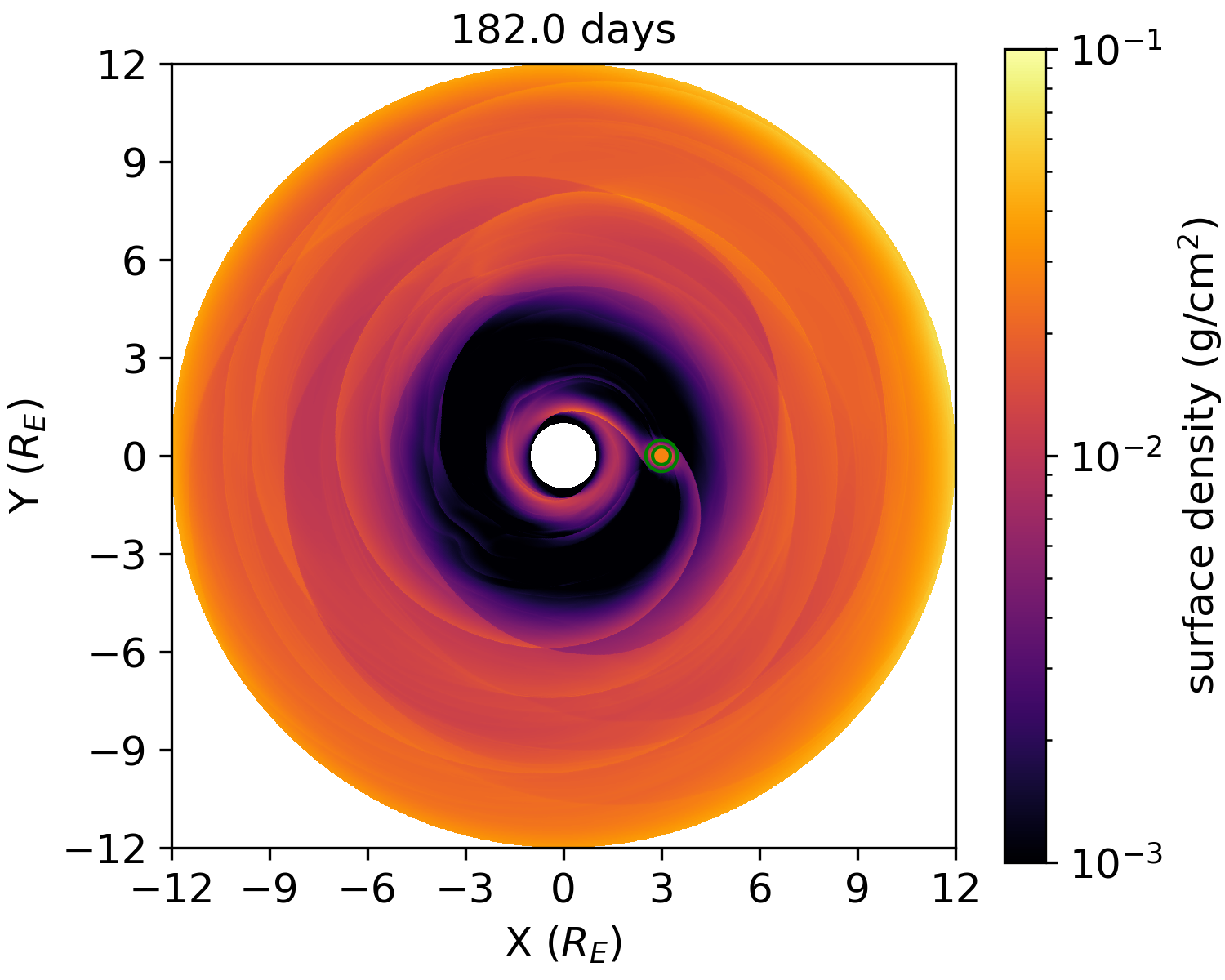}\label{fig:xyminore}}
\quad
\subfloat[]{\includegraphics[width=0.6\columnwidth]{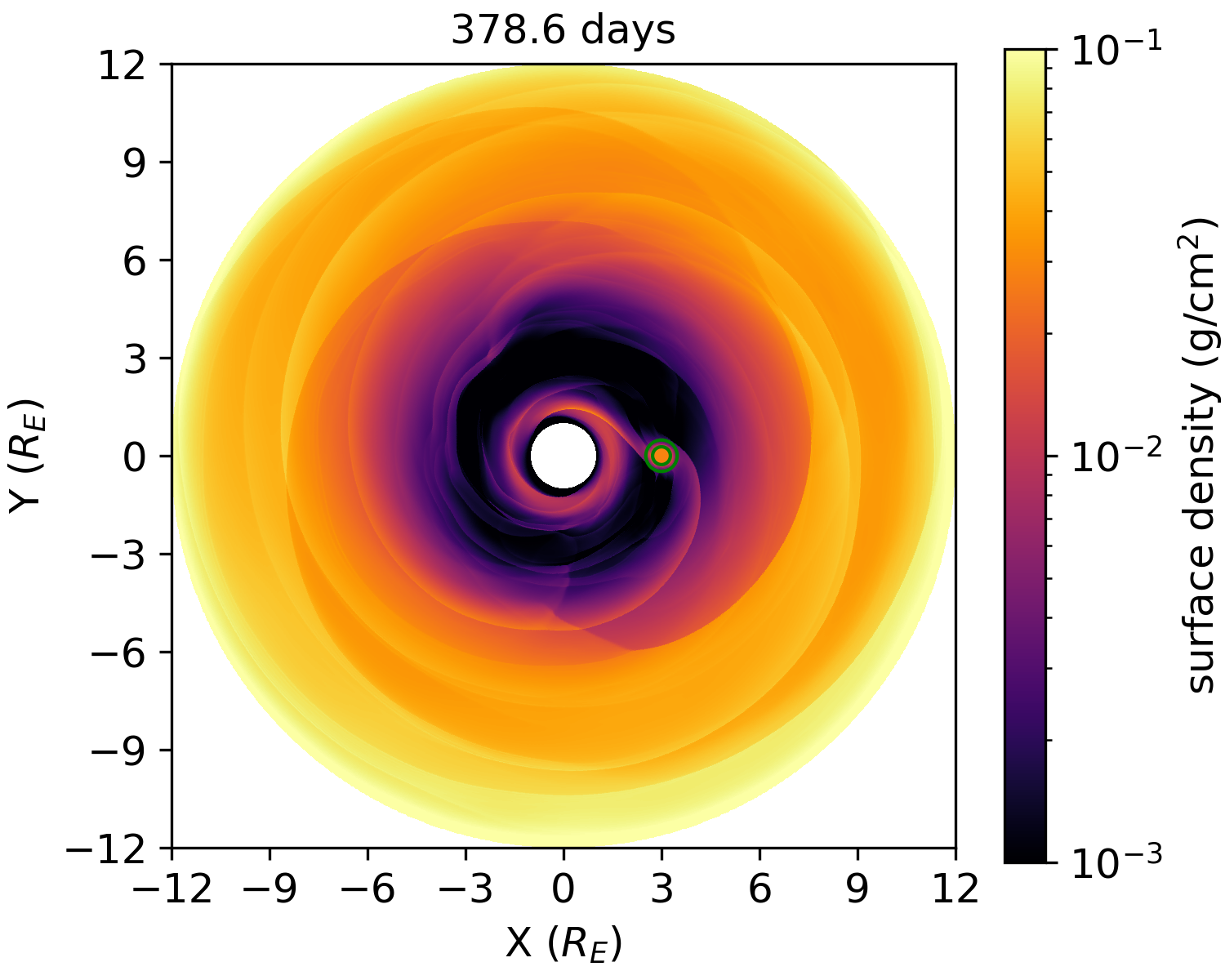}\label{fig:xyminorf}} \\
\centering
\subfloat[]{\includegraphics[width=0.9\columnwidth]{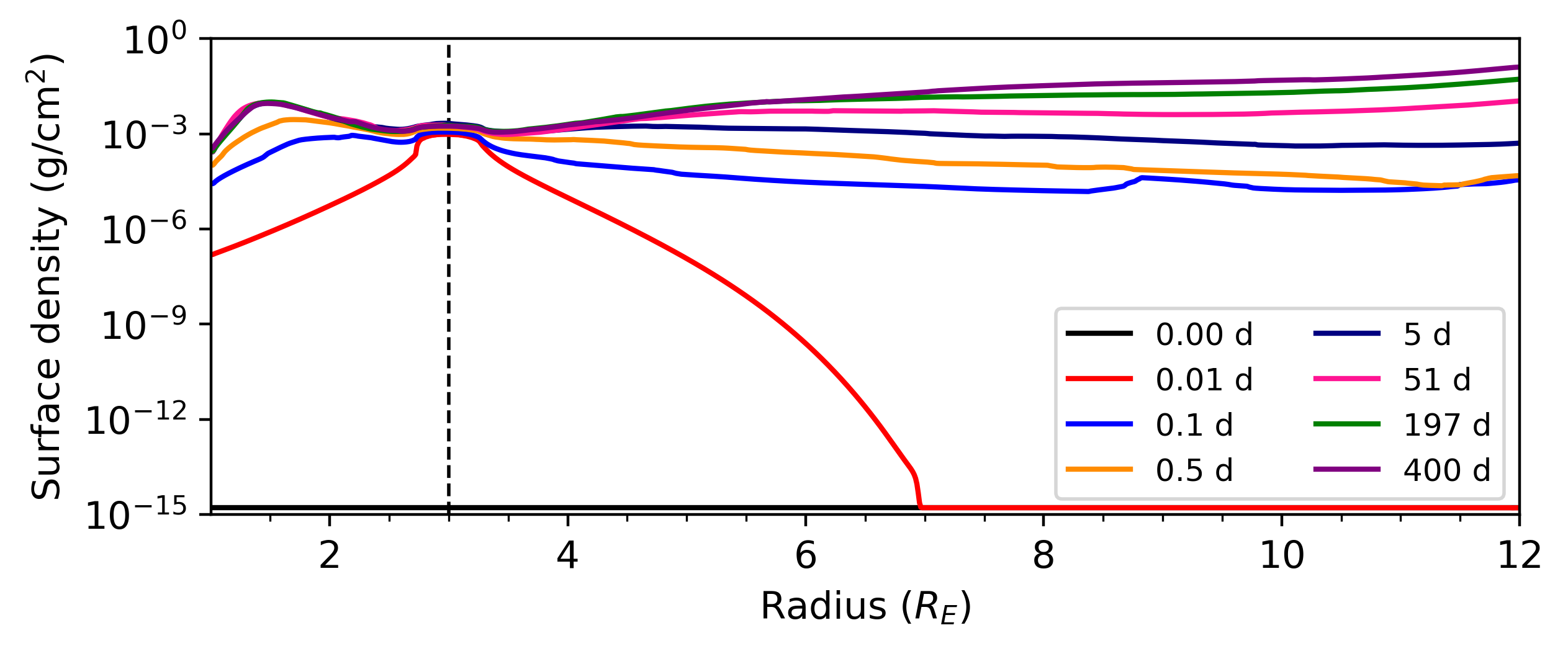}\label{fig:xyminorg}}
\quad
\subfloat[]{\includegraphics[width=0.9\columnwidth]{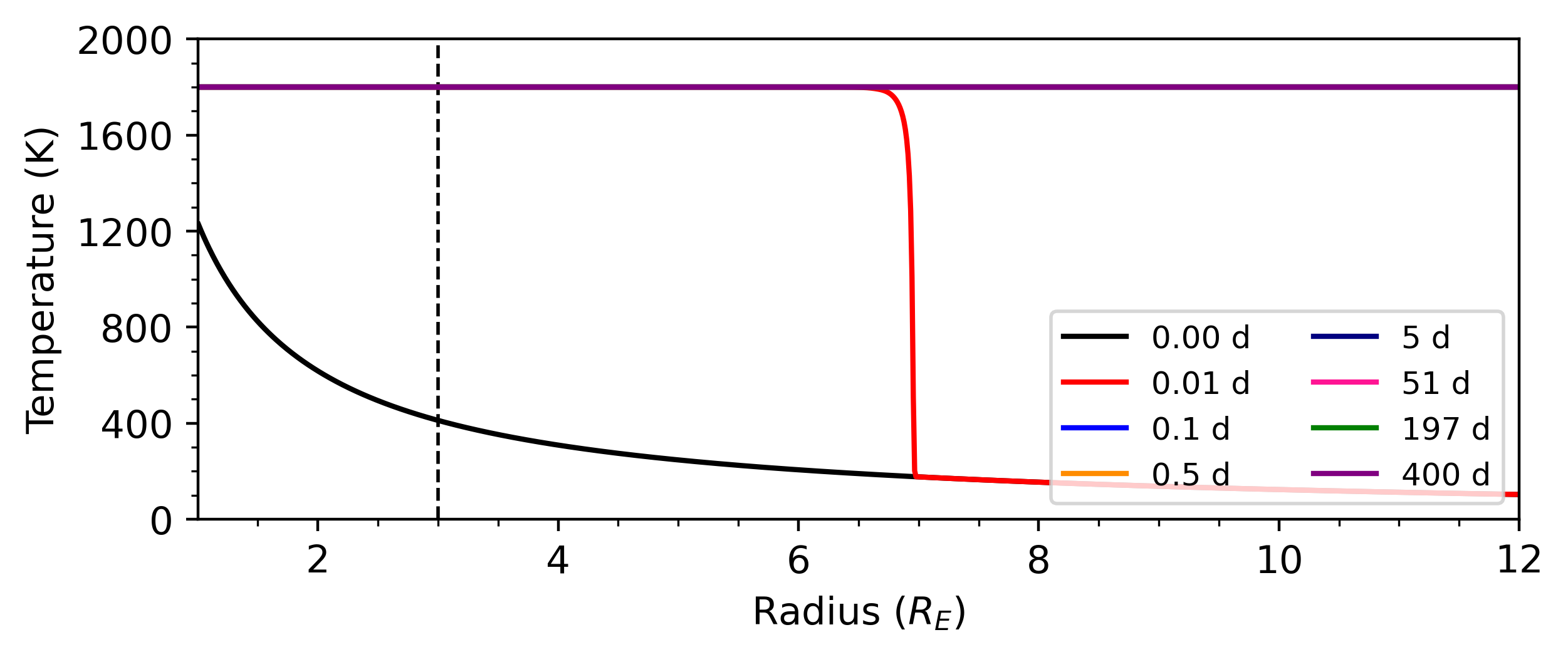}\label{fig:xyminorh}}
\caption{Outcome of the simulation with the molten Moon located at 3~$R_E$ and a surface temperature of 1800~K. Panels (a)-(f) show the surface density of the circum-Earth disk, with green circles indicating the lunar physical radius and the Hill sphere. Panels (g) and (h) show the radial surface density and radial temperature of the disk, respectively. The system is presented in the rotating frame with the Moon.}
\label{fig:xyminor}
\end{figure*}

In this section, we explore the evolution and fate of the gas vapor released from the lunar surface. Initially, we focus on the case with the Moon at 3~$R_E$, position at which lunar aggregates are expected to form \citep{ida2001n,salmon2012lunar,canup2015lunar}. Figure~\ref{fig:xyminor}\footnote{Animations of Figures~\ref{fig:xyminor} and \ref{fig:xymajor} can be found at \url{gusmadeira.github.io/data}} shows the outcome of the simulation with the Moon's surface temperature $T_M$=1800~K. Panels \ref{fig:xyminora}-\ref{fig:xyminorf} show the surface density of the disk at different times, while panels \ref{fig:xyminorg} and \ref{fig:xyminorh} give the radial profile of surface density and temperature, respectively.

For a Moon located at 3~$R_E$, we obtain that the gas vapor is energetic enough to cross the lunar Hill sphere and enter in orbit around the Earth, corroborating the scenario proposed by \cite{charnoz2021tidal}. The gas exits the region gravitationally dominated by the Moon via spiral arms, resulting in the formation of a circum-Earth disk mainly composed of volatile species. The evolution of the disk is similar to that expected for a protoplanetary disk with a large planet immersed in it \citep[e.g.,][]{kley2012planet,morbidelli2016challenges,armitage2019planet}, including the formation of the gap at the lunar location.

The material released from the Moon expands throughout the entire disk, which eventually reaches an isothermal steady state. To understand this process, we must distinguish between the gas already present in the disk and the gas recently released from the Moon. The ejected gas, initially at the same temperature as the Moon, decompresses as it expands, resulting in a local drop in temperature. Simultaneously, this expanding gas heats the colder gas already present in the disk. This process repeats continuously, with the disk being progressively heated by the newly released gas until a steady state is reached, which occurs when the surface density of the disk approaches that of the Moon. At this point, the disk's temperature stabilizes at lunar temperature, and the gas only undergoes local compression or decompression in order to maintain a constant temperature.

In our model, we find that increases in the surface temperature lead to an increase in the amount of material released from the Moon. Nonetheless, the general dynamics of the circum-Earth disk are similar to those shown in Figure~\ref{fig:xyminor}. The main impact of lunar temperature on the structure of the circum-Earth disk is on the depth of the gap, which decreases with temperature. This result is somewhat expected since the disk height is proportional to the square root of temperature, and the proportional relation between the disk height and the depth of the gap is well-documented \citep{ziampras2020importance}.

\begin{figure}[]
\centering
\subfloat[]{\includegraphics[width=0.9\columnwidth]{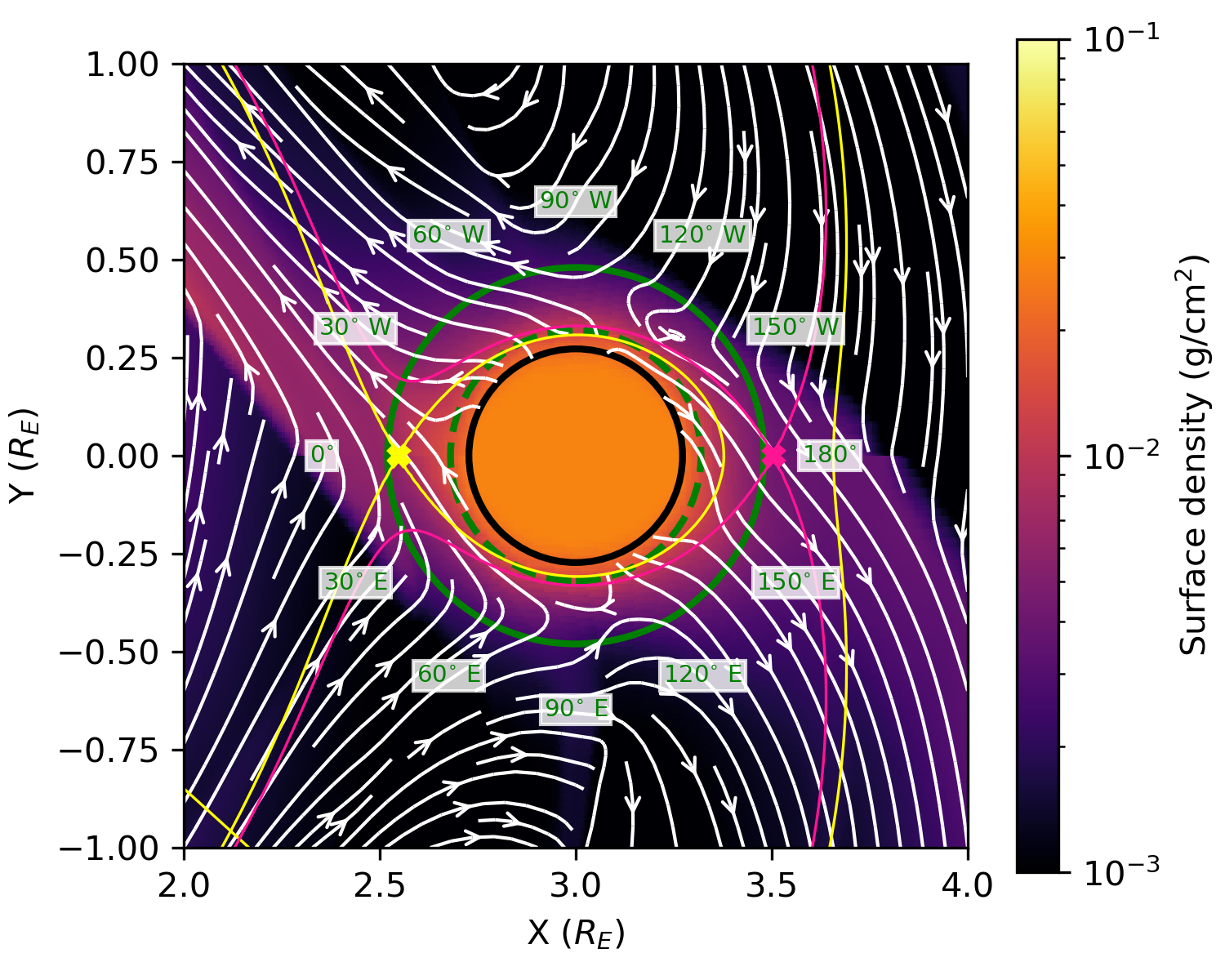} \label{fig:angminora}}\\
\centering
\subfloat[]{\includegraphics[width=0.9\columnwidth]{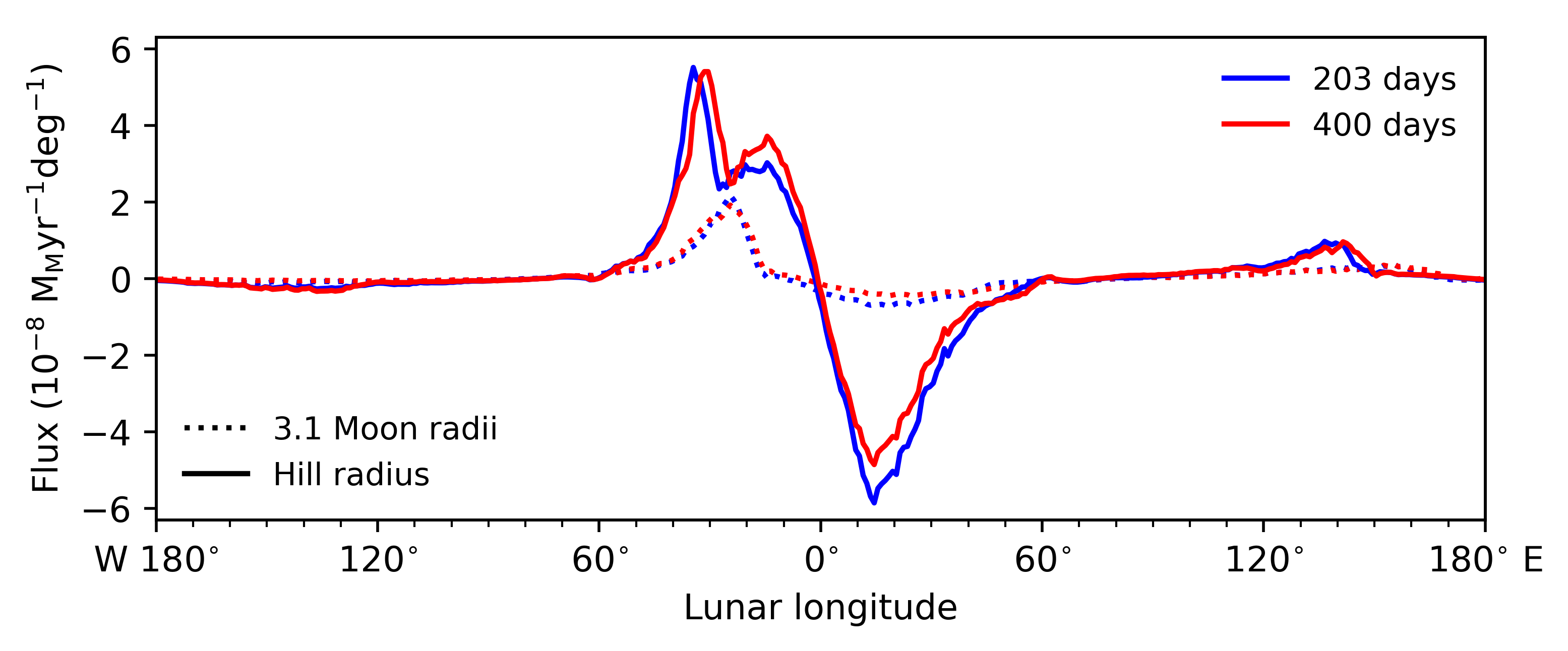} \label{fig:angminorb}}\\
\centering
\subfloat[]{\includegraphics[width=0.9\columnwidth]{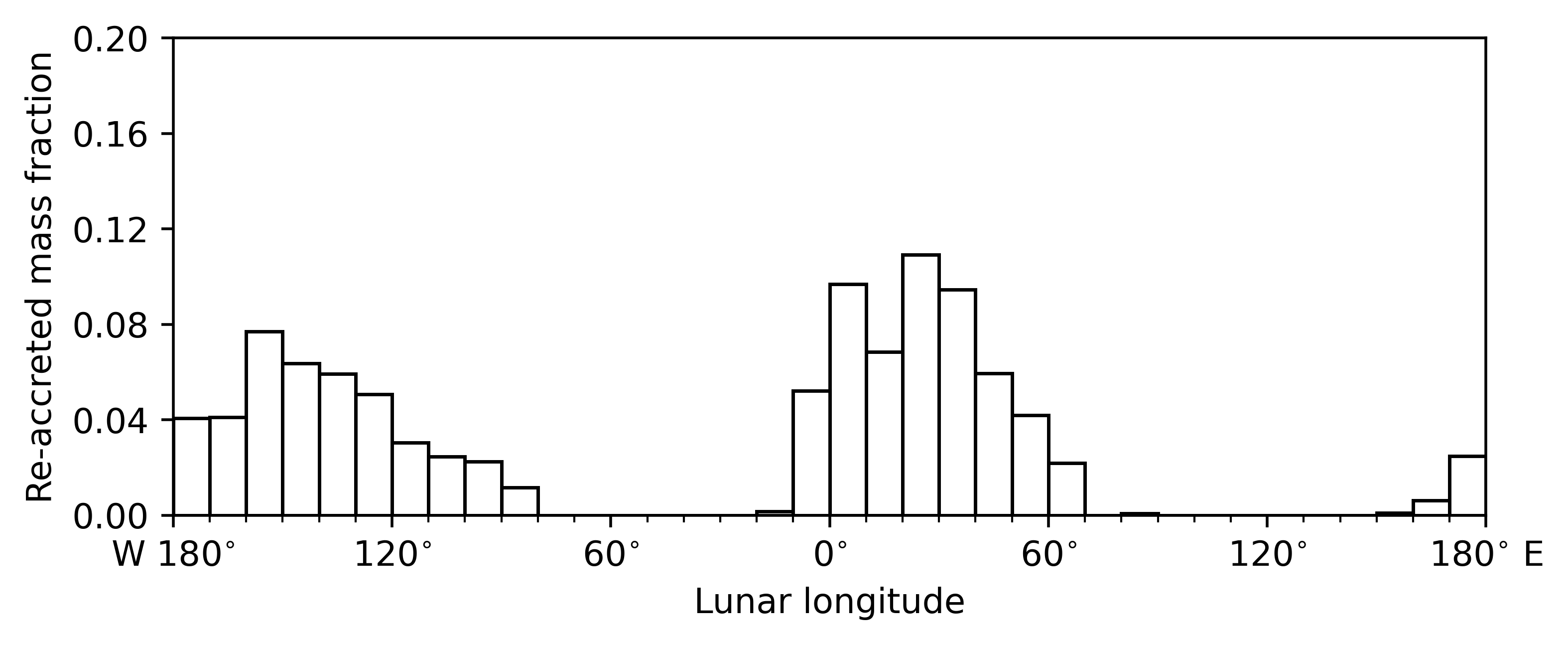} \label{fig:angminorc}}\\
\centering
\subfloat[]{\includegraphics[width=0.9\columnwidth]{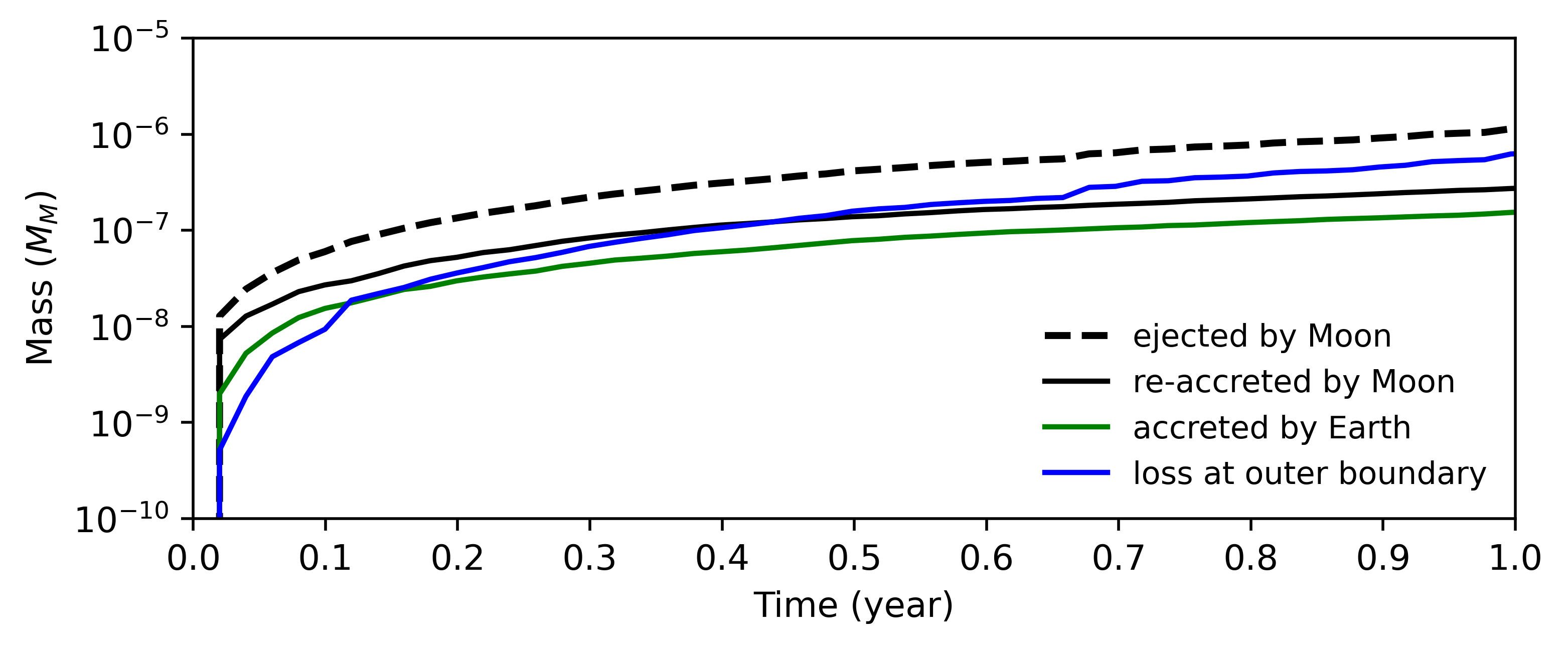} \label{fig:angminord}} \\
\caption{(a) Surface density and motion streamlines in the vicinity of the Moon, where the solid black line indicates the physical radius of the Moon, and the dashed and solid green lines correspond to circumferences with radii equal to the Hill radius $R_{\rm H}$ and 1.2 Moon radii, respectively. The yellow and pink curves represent pure-gravitational equipotential curves related to the L$_1$ and L$_2$ points, respectively. (b) Longitudinal flux density at 1.2 Moon radii (dotted lines) and at the Hill radius (solid lines) at different times, as a function of lunar longitude. (c) Fraction of re-accreted material at different longitudes of the Moon. (d) The amount of material released from the Moon, accreted by Earth, re-accreted by the Moon, and lost at the outer boundary of the disk. The Moon is located at 3~$R_E$ and has a surface temperature of 1800 K.}
\label{fig:angminor}
\end{figure}

Figure~\ref{fig:angminor} illustrates the evolution of the gas material. Panel (a) shows the gas motion streamlines at a selected time, while panel (b) presents the longitudinal flux density at different distances from the Moon, as a function of lunar longitude\footnote{We define zero longitude on the line connecting the Earth and the Moon, on the near side. The west corresponds to the leading side, and the east to the trailing side.}. Panel (c) shows the fraction of re-accreted material as a function of lunar longitude, and panel (d) illustrates the amount of material released from the Moon, re-accreted by the satellite, and lost at the inner and outer boundaries of the disk.

Material is expected to enter orbit around the Earth when crossing the lunar Hill sphere (solid green curve in Fig.~\ref{fig:angminora}), which is theoretically computed using the simplified relation \(R_{\rm H} = a_M(M_M/(3M_E))^{1/3} \approx 0.6a_M/R_E\) in Moon radii. Nonetheless, when the Moon is close to Earth, planetary tides significantly distort the lunar Roche lobe \citep{Kopal1972,murray2000solar}, meaning that a perfectly symmetrical Hill sphere does not accurately represent the region where the Moon's gravity is dominant. This can be seen by the yellow and pink pure-gravitational equipotential curves in Fig.~\ref{fig:angminora}, related to the L$_1$ and L$_2$ points of the system\footnote{The equipotential curves are computed without accounting for the pressure effect of the circum-Earth disk. Pressure effects are known to only slightly change the position of L$_1$ and L$_2$ \citep[Fig. 10,][]{Casoli2009}.}, respectively. To fully understand the circulation of material around the Moon, Figure~\ref{fig:angminora} shows the longitudinal flux density not only at the circular Hill radius but also at 1.2 Moon radii (dotted green curve), approximately corresponding to the closest distance of the Roche lobe to the Moon.

The pressure effect of the gas disk is observed to displace the system's Lagrange points \citep{Casoli2009}, causing L$_1$ and L$_2$ to librate around \(30^{\circ}\) W and \(150^{\circ}\) E, respectively. Although volatile gas is released from the entire cross-section of the Moon, we find that the gas flows out of the Moon's Roche lobe via spiral arms passing through L$_1$ and L$_2$. These spiral arm regions correspond to areas of positive flux in Figure~\ref{fig:angminorb}, with higher flux observed on the near side of the Moon. Volatile re-enter the Roche lobe in regions other than the spiral arms, with a fraction of this material flowing toward the Moon and being re-accreted at roughly the same longitudes where it first penetrates the lobe. The circum-Earth disk material re-enters Moon's Roche lobe mostly after orbiting the Earth for a timescale of a few days.

Over the course of one year of simulation, a total mass of $1.1\times10^{-6}$~$M_M$ is released and escapes the Moon's Roche lobe, with approximately 18\% of the material eventually re-entering this region and being re-accreted by the satellite. Consequently, the net loss of volatile is reduced to $9\times10^{-7}$ $M_M$. Of the material not re-accreted, about 18\% is lost at the inner boundary of the disk -- that is, it is accreted by Earth -- and 72\% is lost at the outer boundary of the disk. The remainder remains in the circum-Earth disk until the end of the 1-year simulation.

\begin{figure*}[]
\centering
\subfloat[]{\includegraphics[width=0.6\columnwidth]{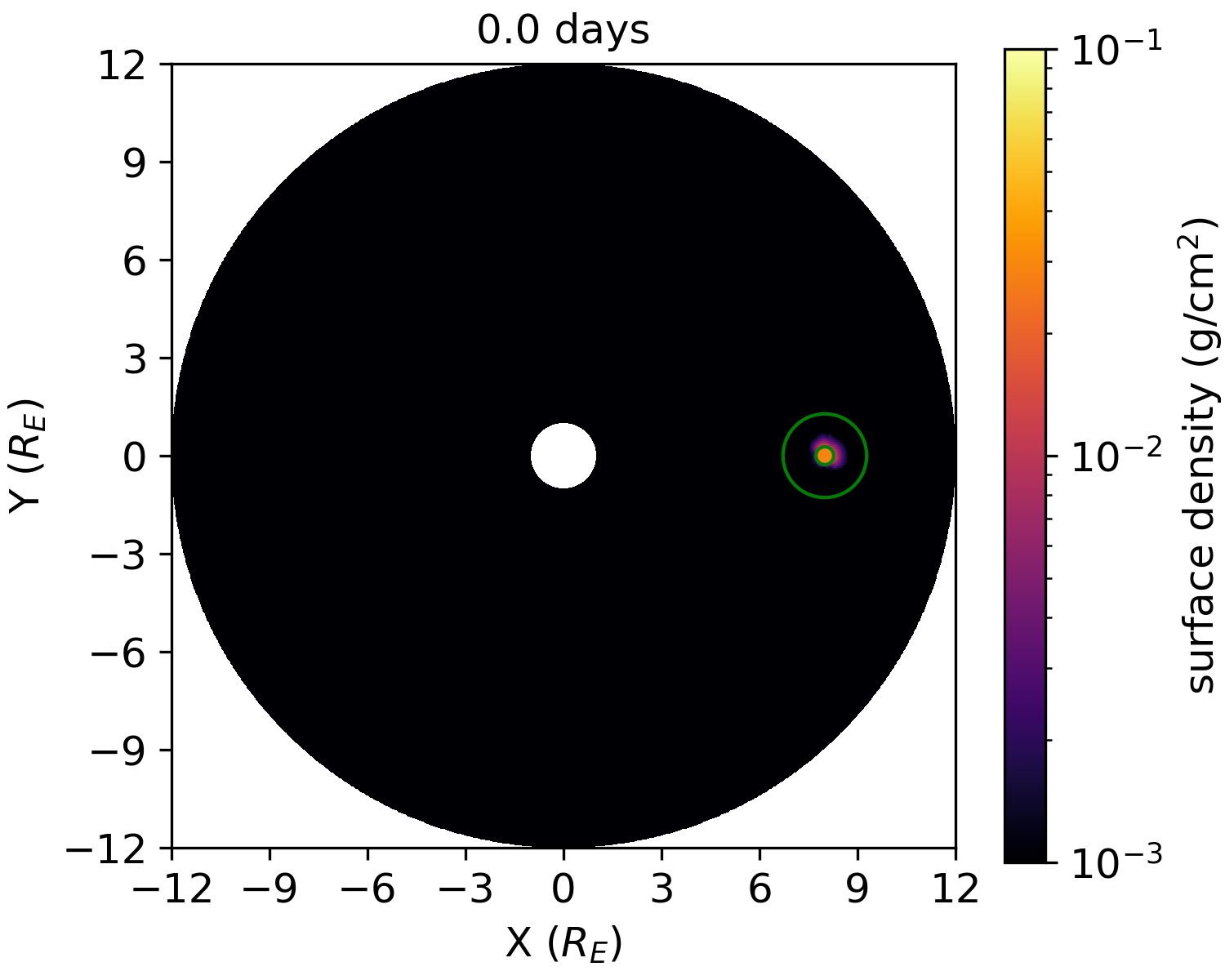}}
\quad
\subfloat[]{\includegraphics[width=0.6\columnwidth]{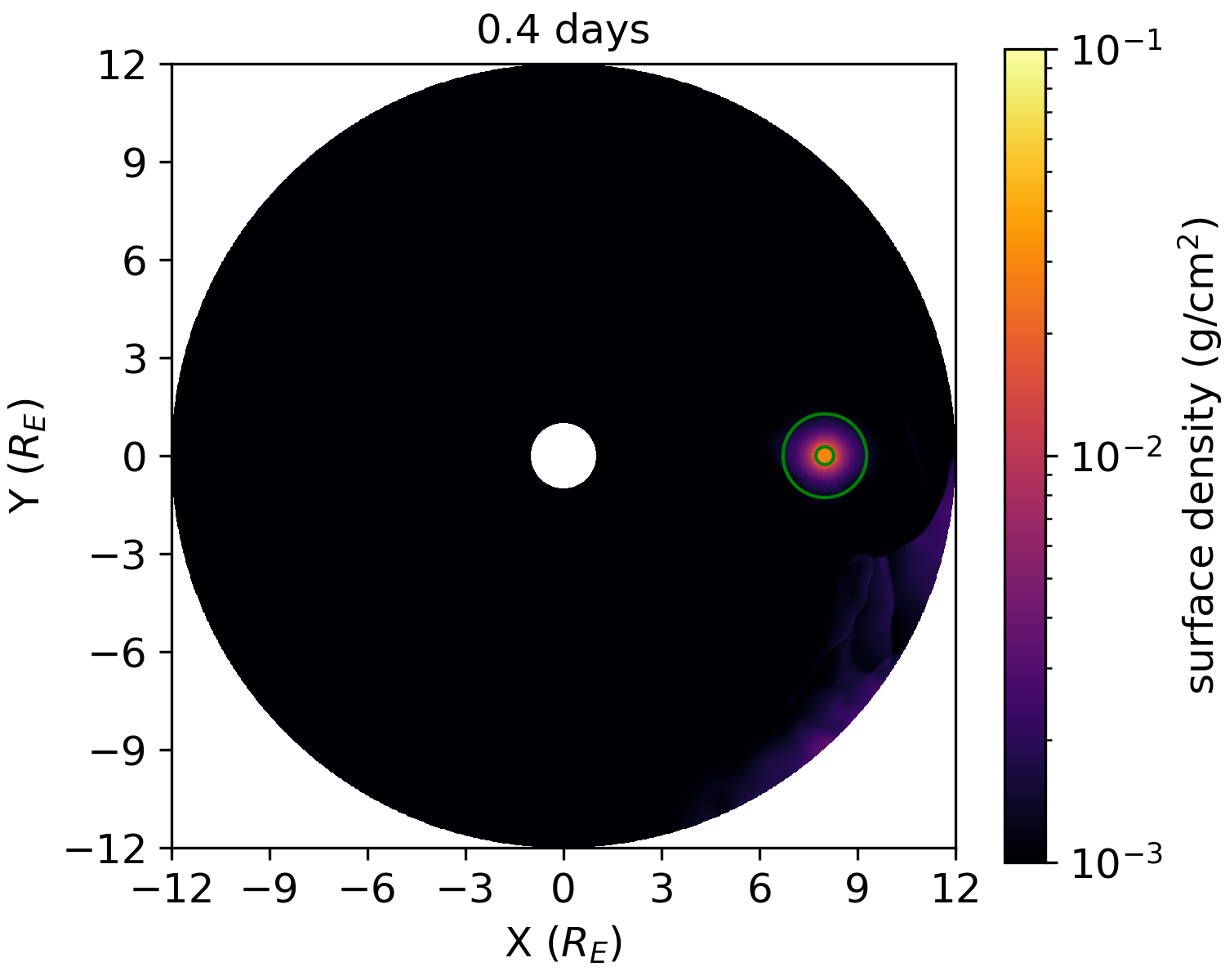}}
\quad
\subfloat[]{\includegraphics[width=0.6\columnwidth]{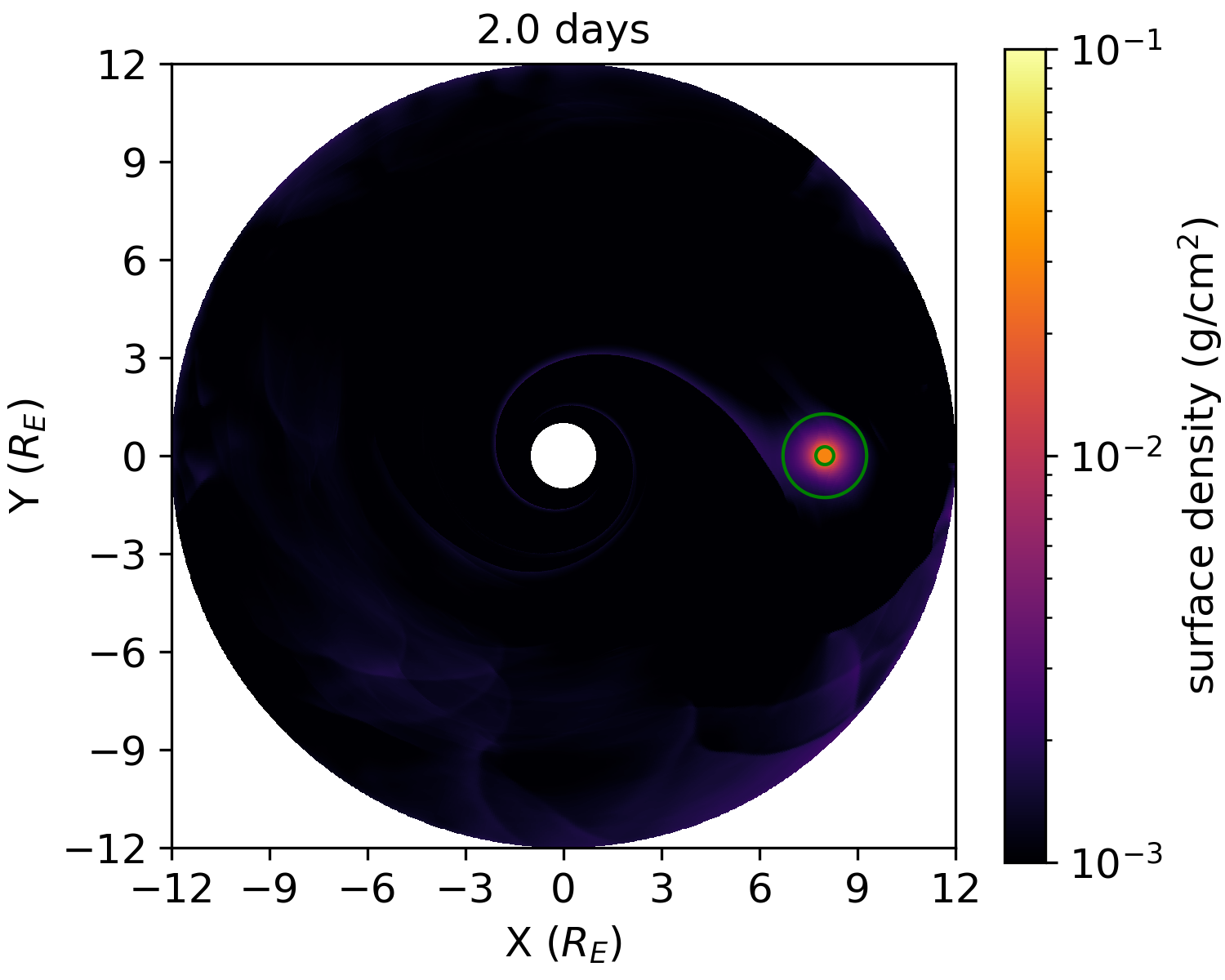}} \\
\centering
\subfloat[]{\includegraphics[width=0.6\columnwidth]{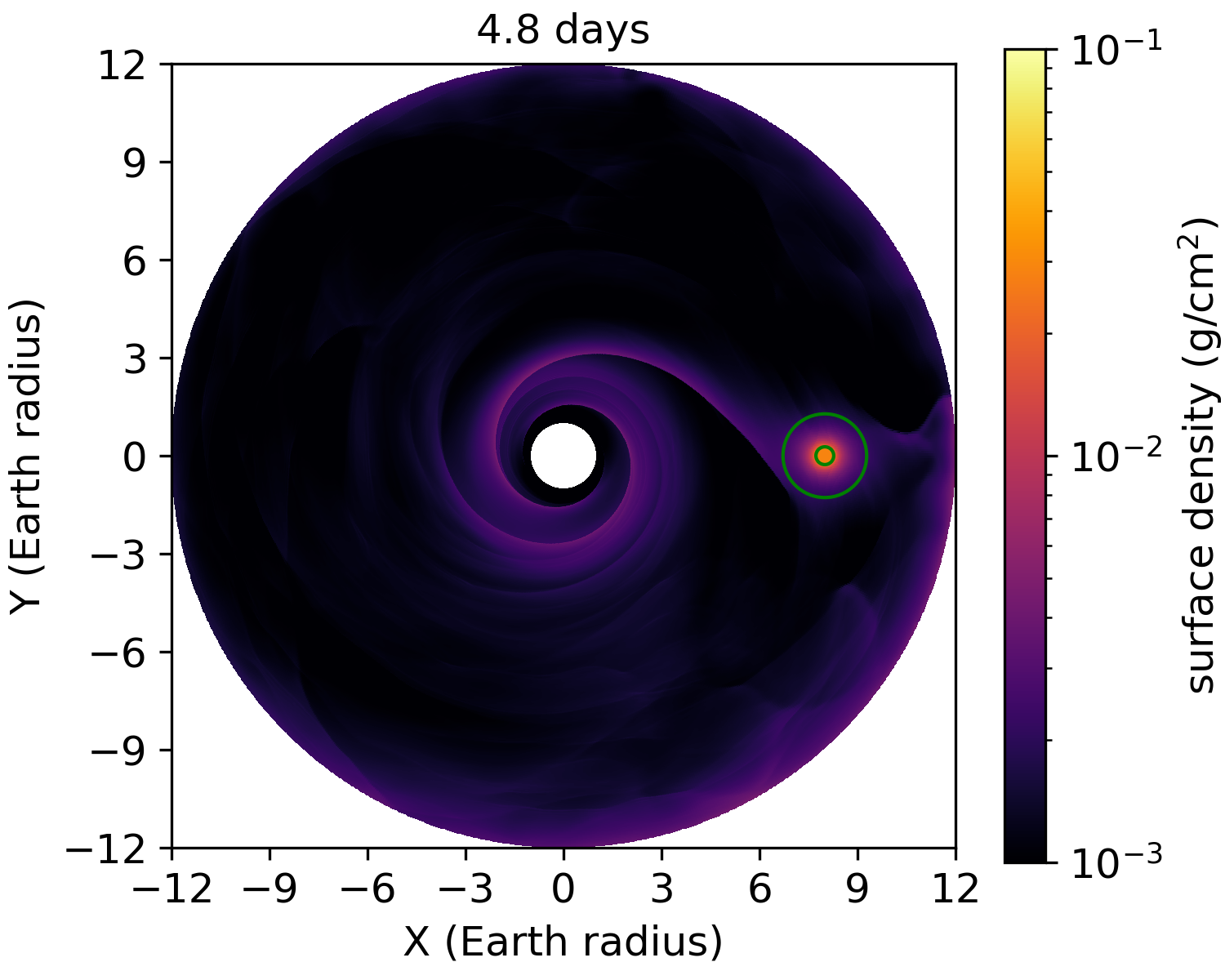}}
\quad
\subfloat[]{\includegraphics[width=0.6\columnwidth]{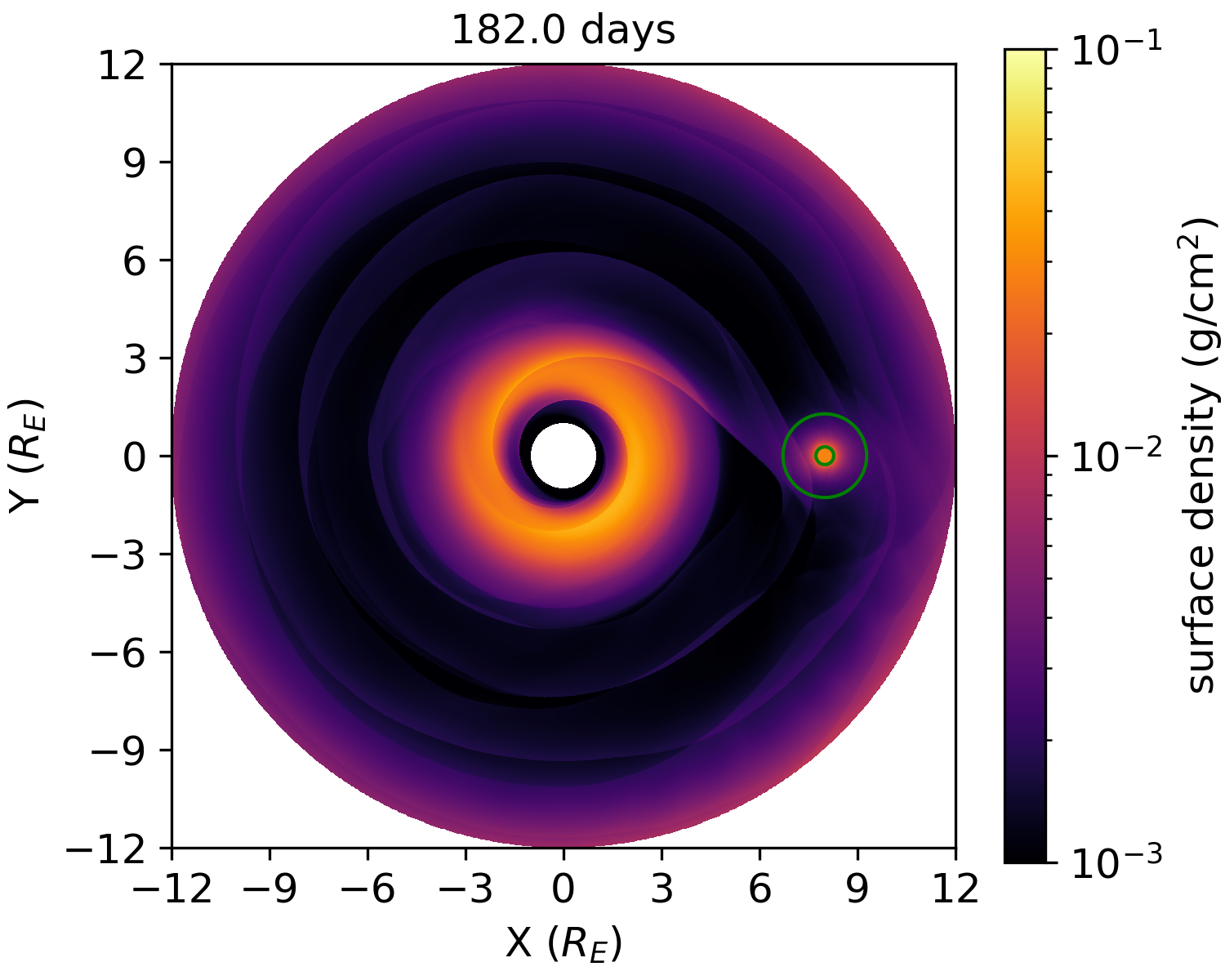}}
\quad
\subfloat[]{\includegraphics[width=0.6\columnwidth]{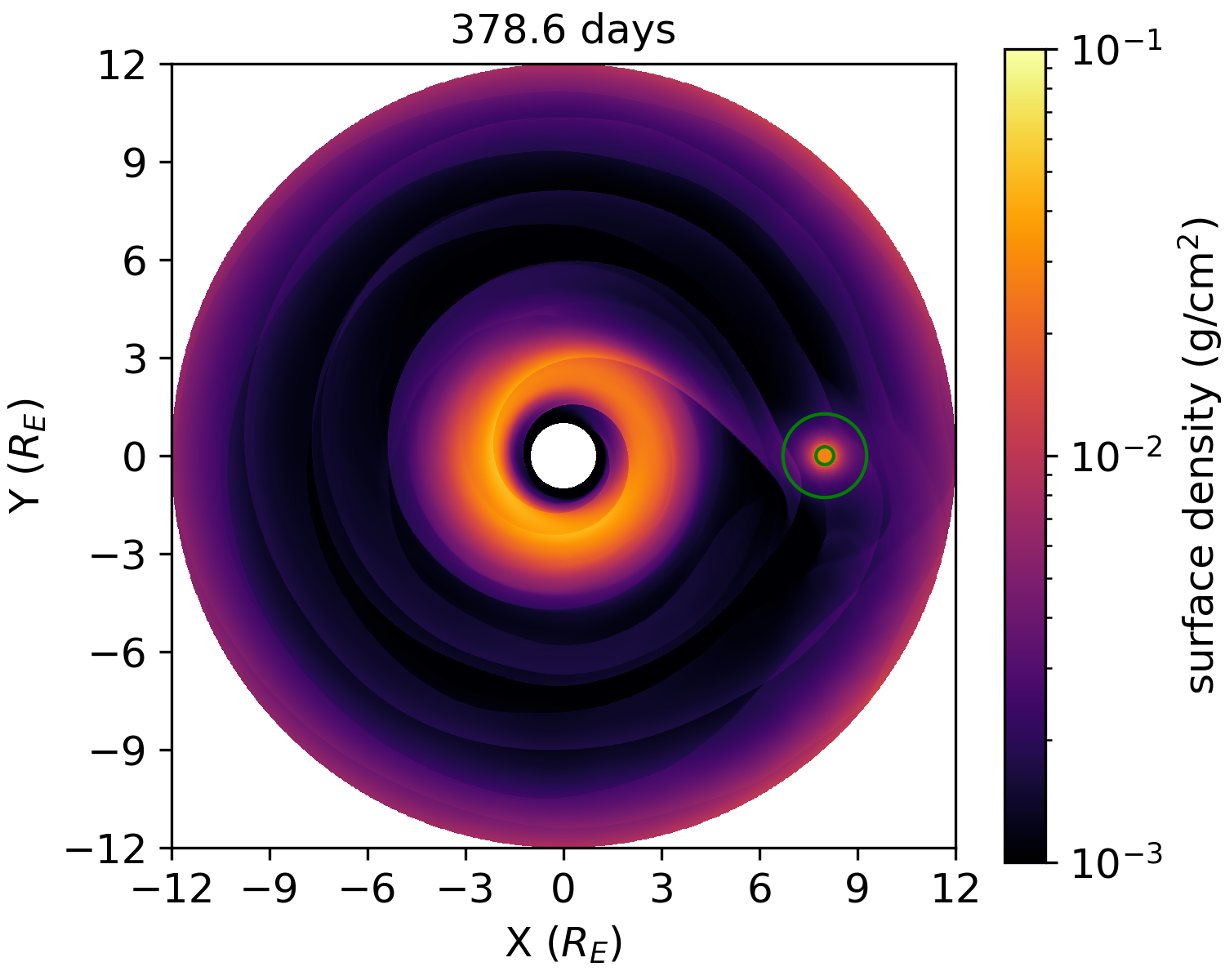}} \\
\centering
\subfloat[]{\includegraphics[width=0.9\columnwidth]{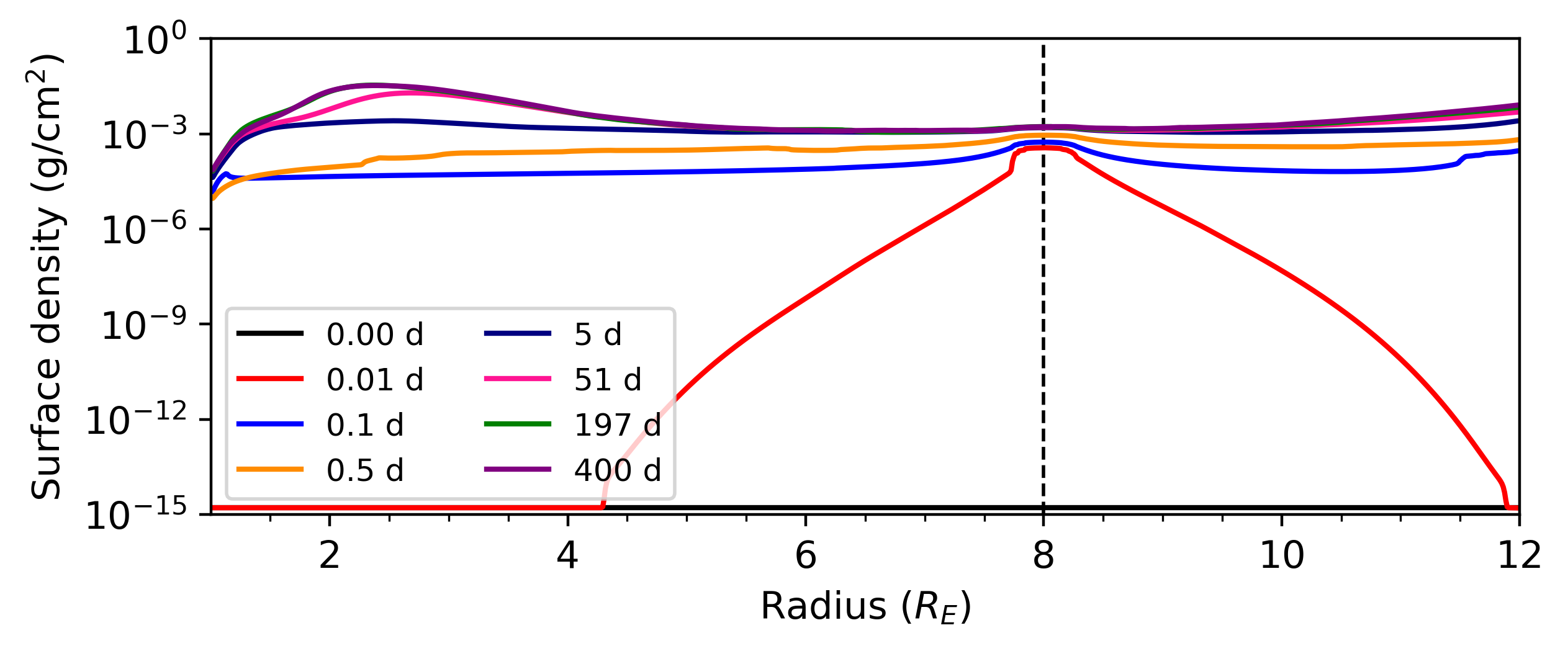}}
\quad
\subfloat[]{\includegraphics[width=0.9\columnwidth]{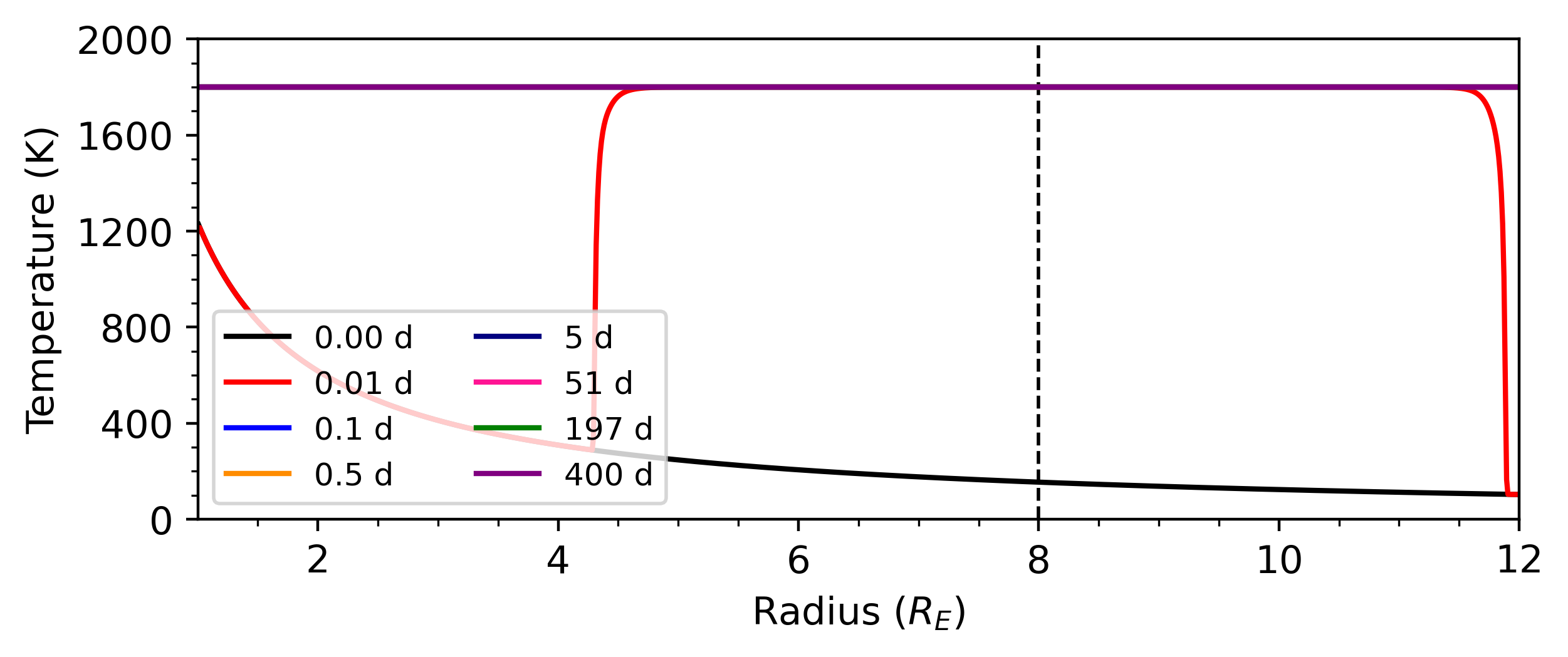}}\\
\caption{Outcome of the simulation with the molten Moon located at~8 $R_E$ with a surface temperature of 1800~K. The surface density of the circum-Earth disk is shown in panels (a) to (f) for different times. The system is in the rotating frame with Moon. The green circles represent the lunar physical radius and the Hill sphere. Panels (g) and (h) show the radial surface density and disk temperature, respectively.}
\label{fig:xymajor}
\end{figure*}
\begin{figure}[]
\centering
\subfloat[]{\includegraphics[width=0.9\columnwidth]{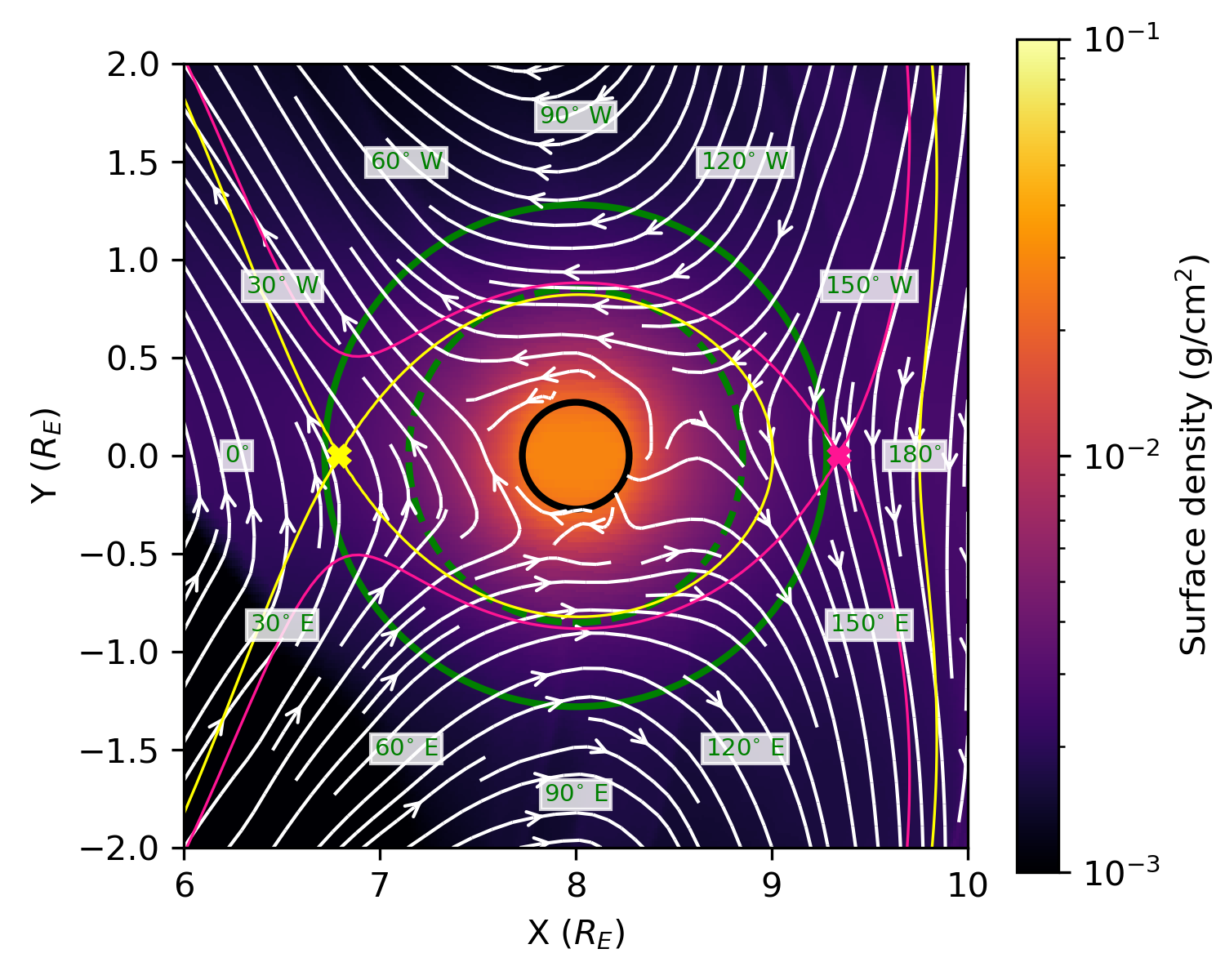}} \\
\centering
\subfloat[]{\includegraphics[width=0.9\columnwidth]{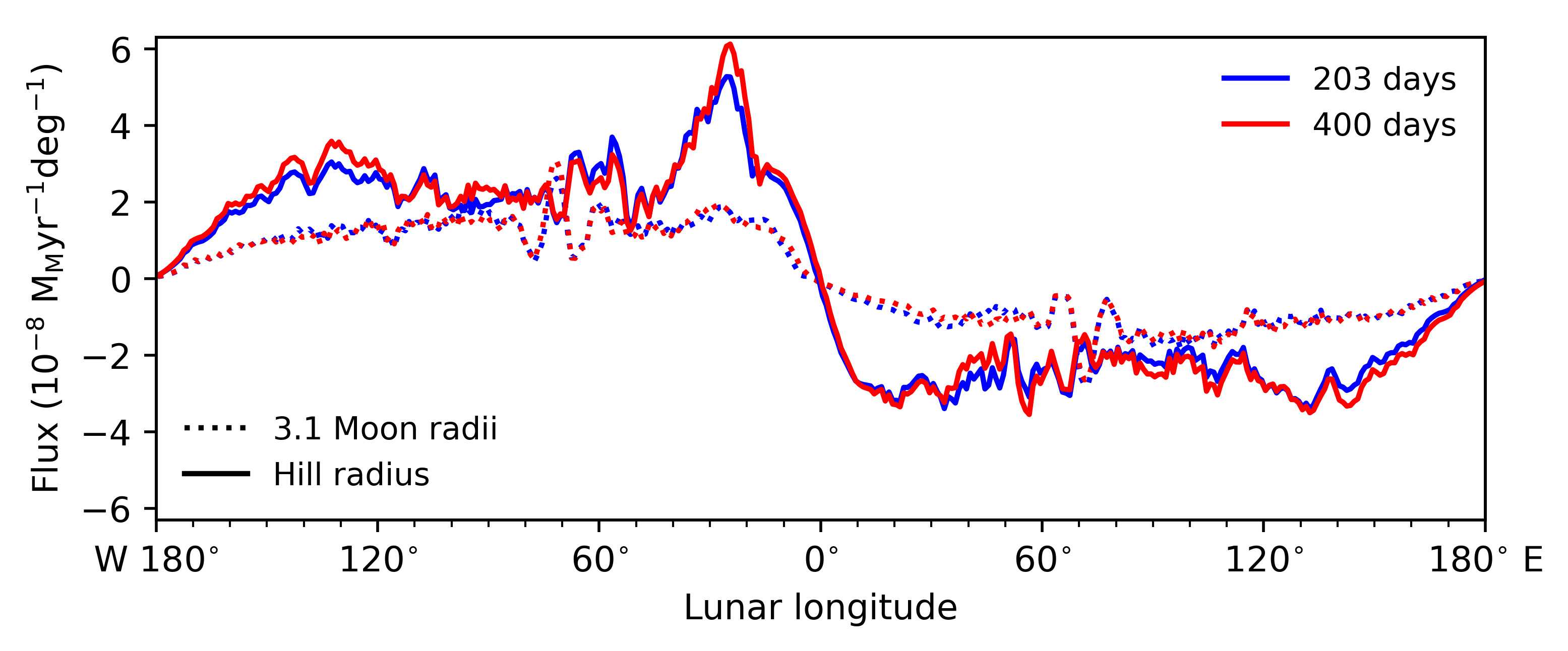}\label{fig:angmajorb}} \\
\centering
\subfloat[]{\includegraphics[width=0.9\columnwidth]{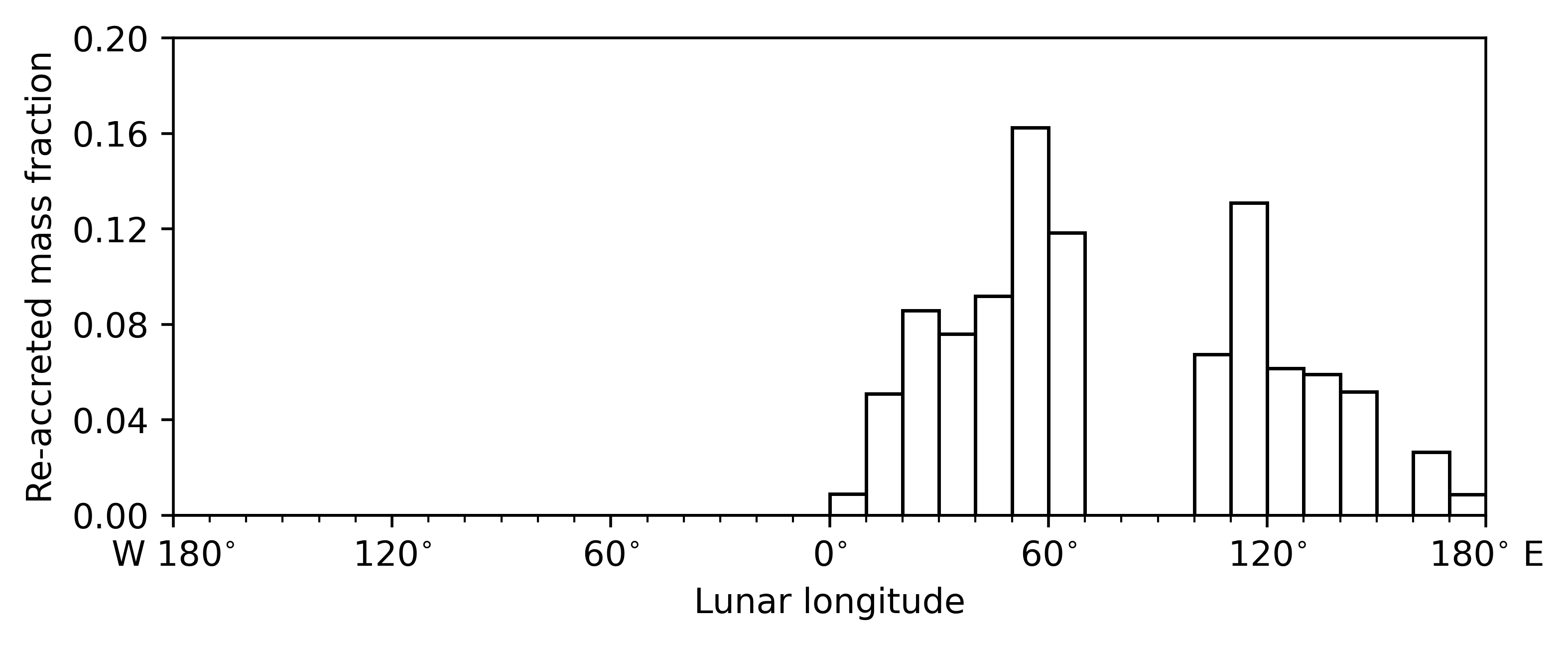}\label{fig:angmajorc}}\\
\centering
\subfloat[]{\includegraphics[width=0.9\columnwidth]{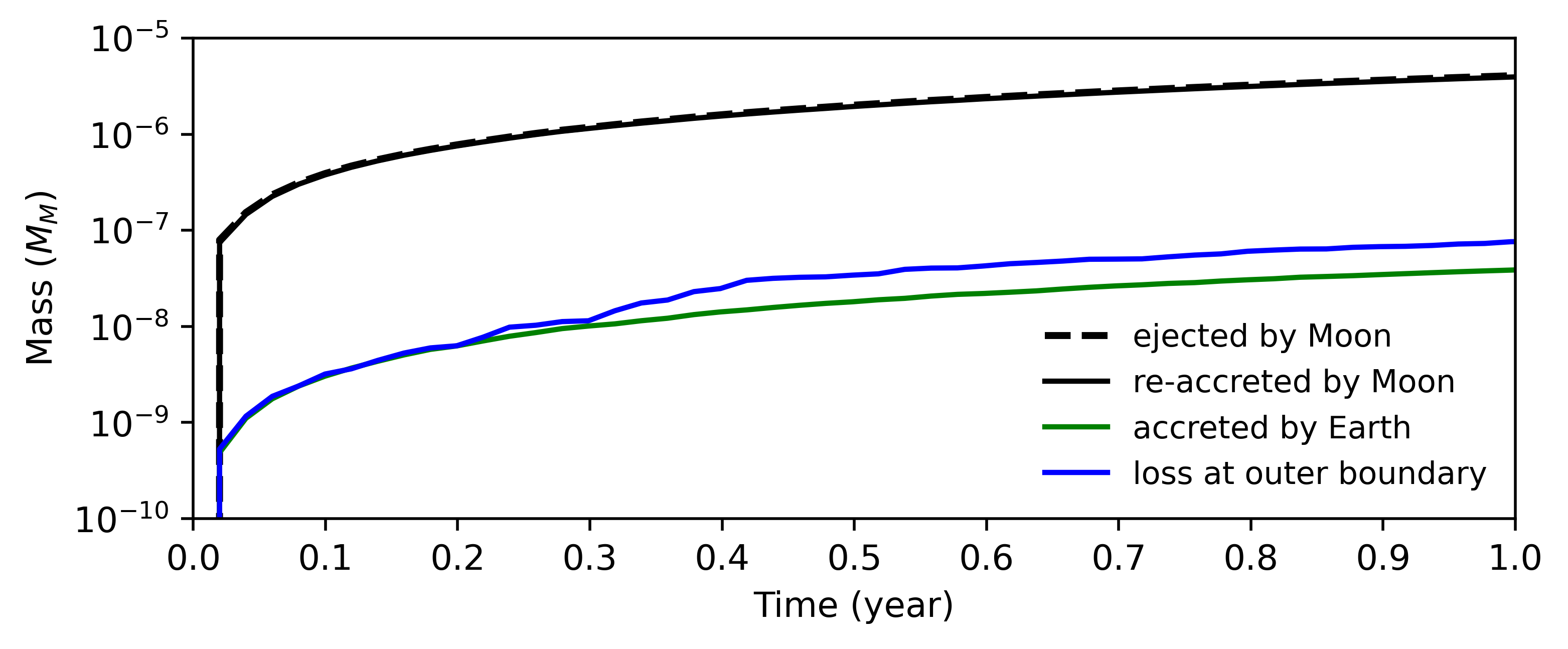}} \\
\caption{(a) Surface density and motion streamlines in the vicinity of the Moon, initially with $a_M=8$ $R_E$ and $T_M=1800$ K. The solid black line indicates the physical radius of the Moon, and the dashed and solid green lines correspond to circumferences with radii equal to the Hill radius and 3.1 Moon radii, respectively. The yellow and pink curves represent equipotential curves related to the L$_1$ and L$_2$ points, respectively. (b) Longitudinal flux density at 3.1 Moon radii (dotted lines) and at the Hill radius (solid lines) at different times. (c) Fraction of the total re-accreted mass as a function of lunar longitude. (d) Amount of material released from the Moon, accreted by Earth, re-accreted by the Moon, and lost at the outer boundary of the disk.}
\label{fig:angmajor}
\end{figure}

Figures~\ref{fig:xymajor} and \ref{fig:angmajor} show the same panels as Figures~\ref{fig:xyminor} and \ref{fig:angminor}, but for the Moon initially at 8~$R_E$. In this case, the theoretical Hill radius corresponds to almost 5 times the Moon's physical radius, an obvious consequence of the new lunar position. The total amount of material released from the Moon is greater than that obtained for $a_M=3~R_E$ ($4.1\times10^{-6}~M_M$), however, only 5\% of the material is actually put into orbit around Earth. Because of this, a much lighter disk of volatile is formed when the Moon is further away from the planet. 

For $a_M=3$ $R_E$, the surface density of the disk reaches peak values of 0.2~g/cm$^2$, while for $a_M=8$~$R_E$, it is always lower than 0.05~g/cm$^2$. For the latter case, it is observed that 40\% of the disk material penetrates the Moon's Roche lobe after few dozen lunar orbits and is re-accreted by the satellite, which implies a small net loss of volatiles in the Moon. The vast majority of the material released from the Moon (95\% of the total) never crosses the lunar Roche lobe and remains in the lunar extended atmosphere, cycling around the satellite for a few hours before being re-accreted. This material tends to move away from the satellite on its leading side and approach it on the trailing side. In fact, in all our simulations with \(a_M > 3.5\)~\(R_E\), volatiles are preferentially re-accreted on the trailing side of the Moon (Fig.~\ref{fig:angmajorc}), likely due to the satellite's orbital motion. In the next section, we analyze the fluxes of gas in the system.

\subsection{Fluxes of ejection and accretion of volatile} \label{sec:fluxes}
\begin{figure}[]
\centering
\subfloat[]{\includegraphics[width=\columnwidth]{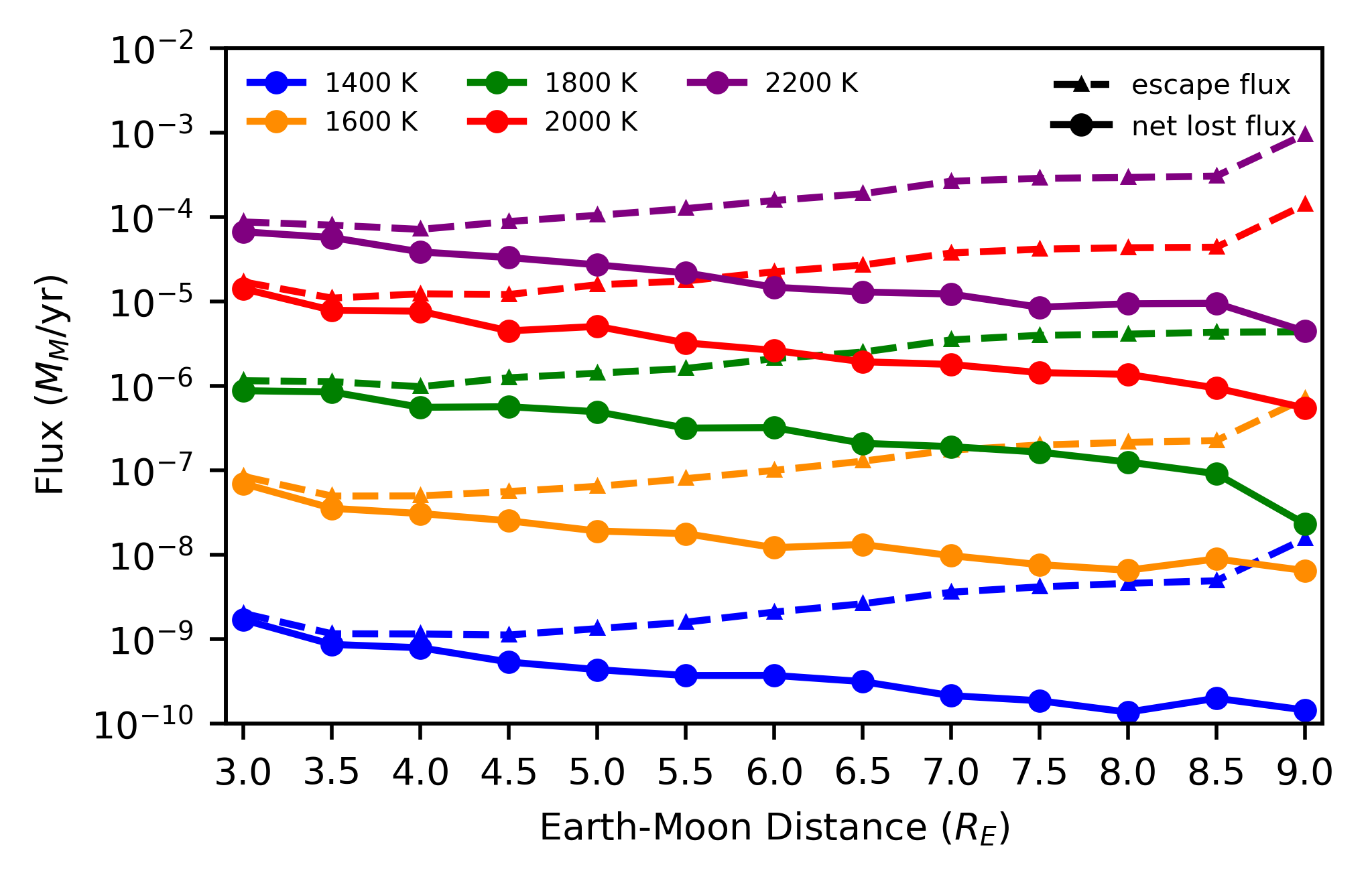}\label{fig:magmaoceana}}\\
\subfloat[]{\includegraphics[width=\columnwidth]{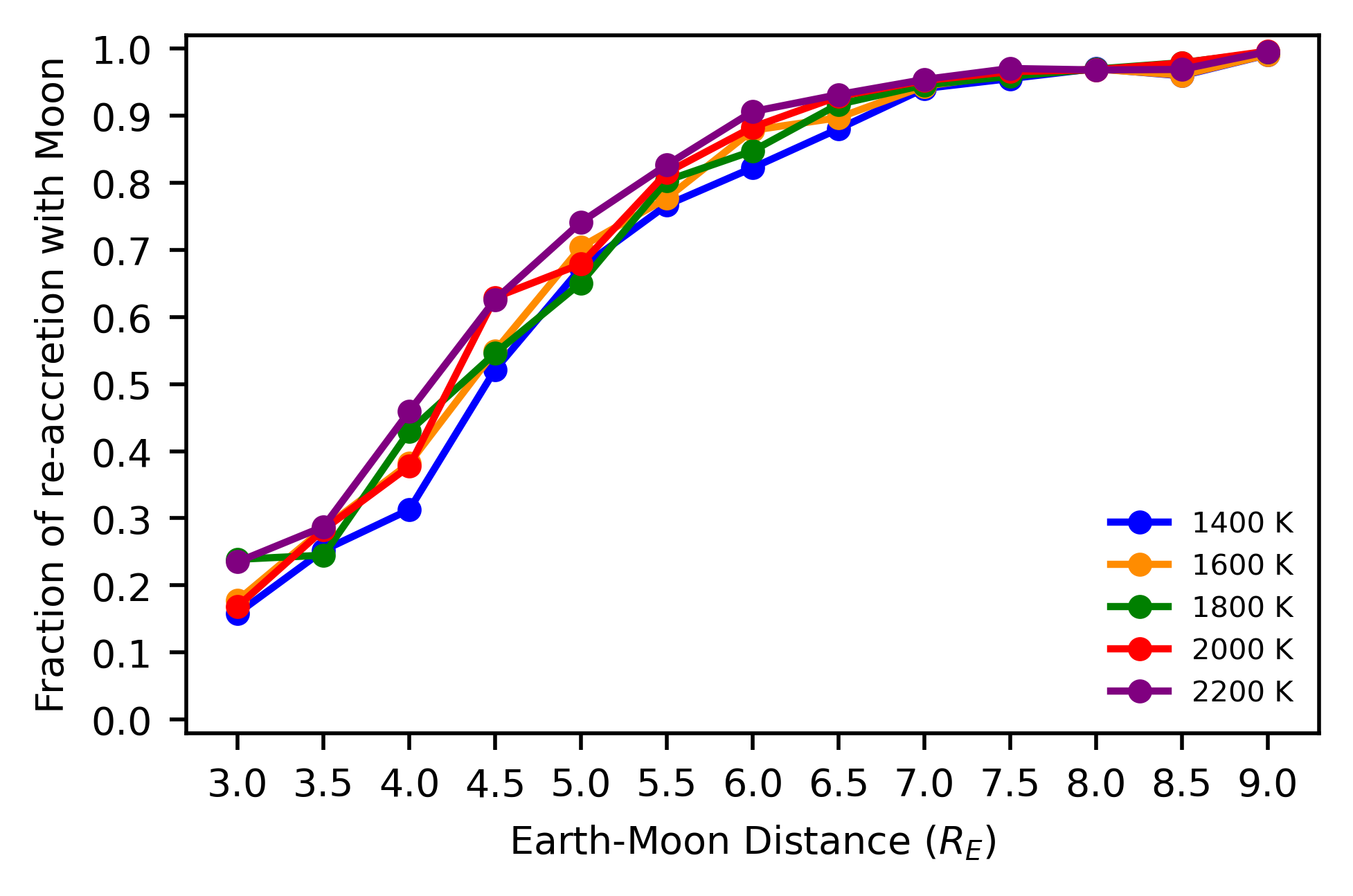}\label{fig:magmaoceanb}}\\
\subfloat[]{\includegraphics[width=\columnwidth]{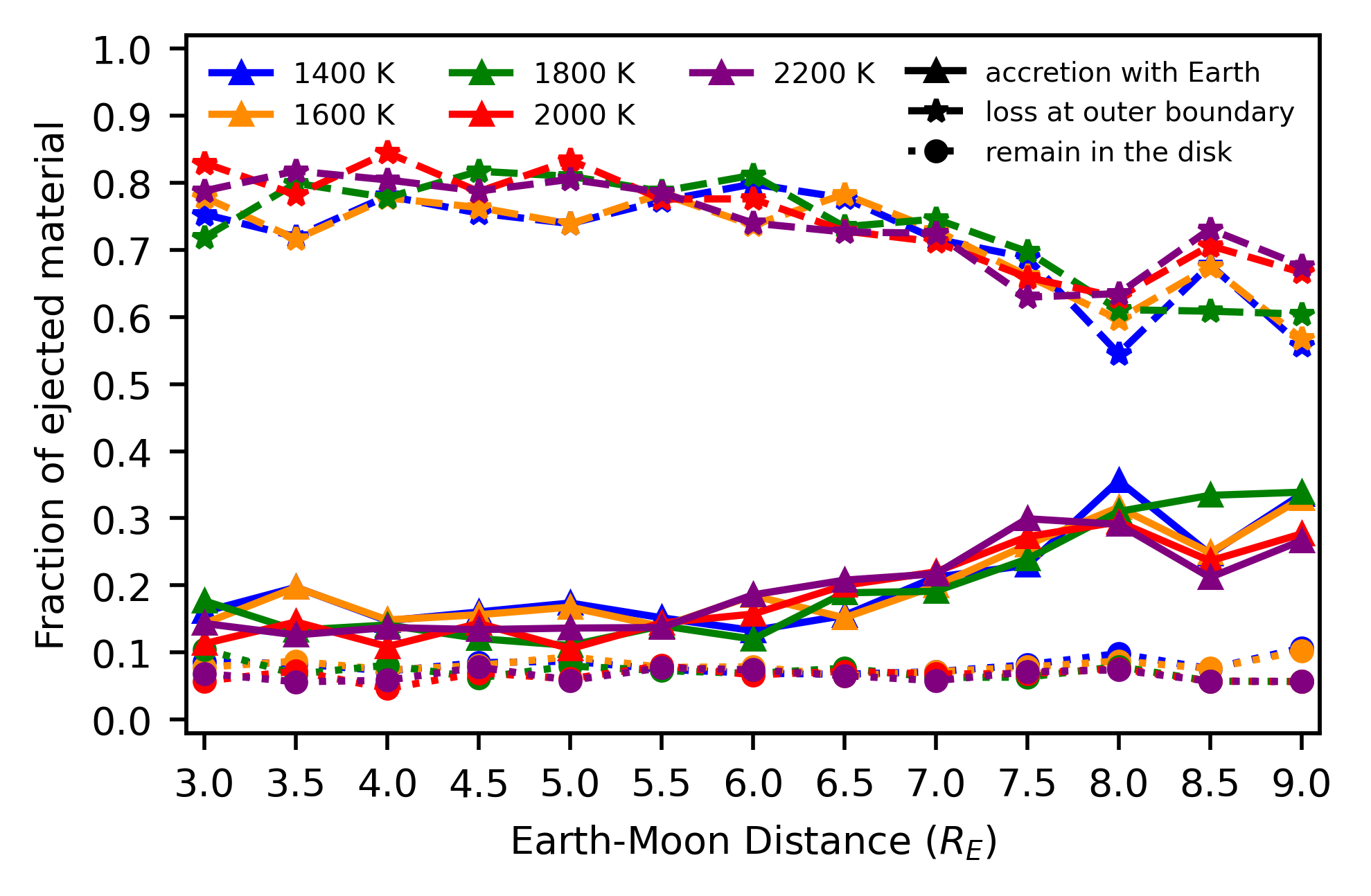}\label{fig:magmaoceanc}}\\
\caption{Panel (a) shows the flux of gas vapor ejected from the Moon (dashed lines) and the flux of material effectively lost by the satellite (solid lines). Panel (b) presents the fraction of the ejected material that is re-accreted by the satellite. Panel (c) shows the fraction of material effectively lost from the Moon that is accreted by Earth (solid lines), lost at the outer boundary (dashed lines), and that remains in the circum-Earth disk (dotted lines) at the end of the simulation. All plots are presented as functions of the Earth-Moon distance.}
\label{fig:magmaocean}
\end{figure}

Figure~\ref{fig:magmaocean} gives the fluxes and fate of material ejected from the Moon in our 1-year simulations, plotted as functions of Earth-Moon distance and surface temperature. As seen in Table~\ref{tab:bse}, the pressure is highly sensitive to surface temperature. Consequently, the lunar surface density -- and therefore the ejection flux -- increases with temperature (dashed lines in Fig.\ref{fig:magmaoceana}), despite the inverse relation between $\Sigma_M$ and $T_M$ in Equation~\ref{eq:sigma}. For $T_M$=1400~K, the ejection fluxes are of the order of $10^{-9}$~$M_M$/year, while for $T_M$=2200~K, they increase to $10^{-4}$ $M_M$/year. It is noteworthy that although the surface density does not depend on the Moon's semi-major axis, the height of the background disk does, which influences the ejection fluxes. This explains the increase in ejection flux with distance.

Nevertheless, the ejection flux is not representative of the material actually lost by the Moon, since some of it returns to the satellite. As discussed earlier, an increasing fraction of material is reaccreted by the Moon without ever crossing the Roche lobe as the Moon's semi-major axis expands. In addition, some of the material actually put in orbit around the Earth is re-accreted by the satellite after a few orbital periods. The re-accretion fraction (Fig.~\ref{fig:magmaoceanb}), combines the re-accretion in these both scenarios, and shows a weak dependence on the surface temperature. For $a_M$=3~$R_E$, $\sim$15-20\% of the ejected material is reaccreted by the Moon, while for $a_M$=9~$R_E$, only about 1\% of the ejected material does not return to the satellite.

The net loss flux from the Moon is given by the solid lines in Fig.~\ref{fig:magmaocean}. A direct comparison of the fluxes obtained in this study with those of the dry model of \cite{charnoz2021tidal} (their Figure 7) is not feasible due to the different assumptions about the vapor gas. Specifically, \cite{charnoz2021tidal} assumes constant molar mass (35 g/mol) and adiabatic index ($\approx$1.5) while assuming the pressure given by the adiabatic equation of state. Our model, on the other hand, assumes these quantities as functions of the lunar temperature. As a result, we obtain a much smoother decrease in the net loss flux with the distance, compared to the estimates of \cite{charnoz2021tidal}. This suggests that the tidally-assisted hydrodynamic escape may be more efficient than previously envisioned, especially considering that our results take into account the material re-accreted by the Moon -- a factor not considered in \cite{charnoz2021tidal}.

The figure~\ref{fig:magmaoceanc} presents the fate of the material not re-accreted by the Moon, showing that the lunar temperature has a small impact on the fraction of material lost at the edges of the disk. In general, about 10\% of the material remains in the disk at the end of the simulation, while most of it is lost at the outer edge, which we interpret as a tendency of the material to be released into the interplanetary environment. Some of the gas is also accreted by the Earth, indicating that the tidally assisted hydrodynamic escape not only devolatilizes the lunar mantle, but also contributes to the enrichment of the Earth with volatile materials. The fraction accreted by the Earth increases with distance, reaching approximately 10\% of the material not recreated by the Moon for $a_M$=3~$R_E$, and reaching $\sim$35\% for $a_M$=9~$R_E$. However, it is important to note that in absolute terms, the amount of material accreted by Earth when the Moon is at 3 $R_E$ is greater than it is at larger distances, due to the greater amount of gas injected in the circum-Earth disk.

\section{Moon's orbital expansion and composition} \label{sec:finalcomposition}
\begin{figure}[]
\centering
\includegraphics[width=\columnwidth]{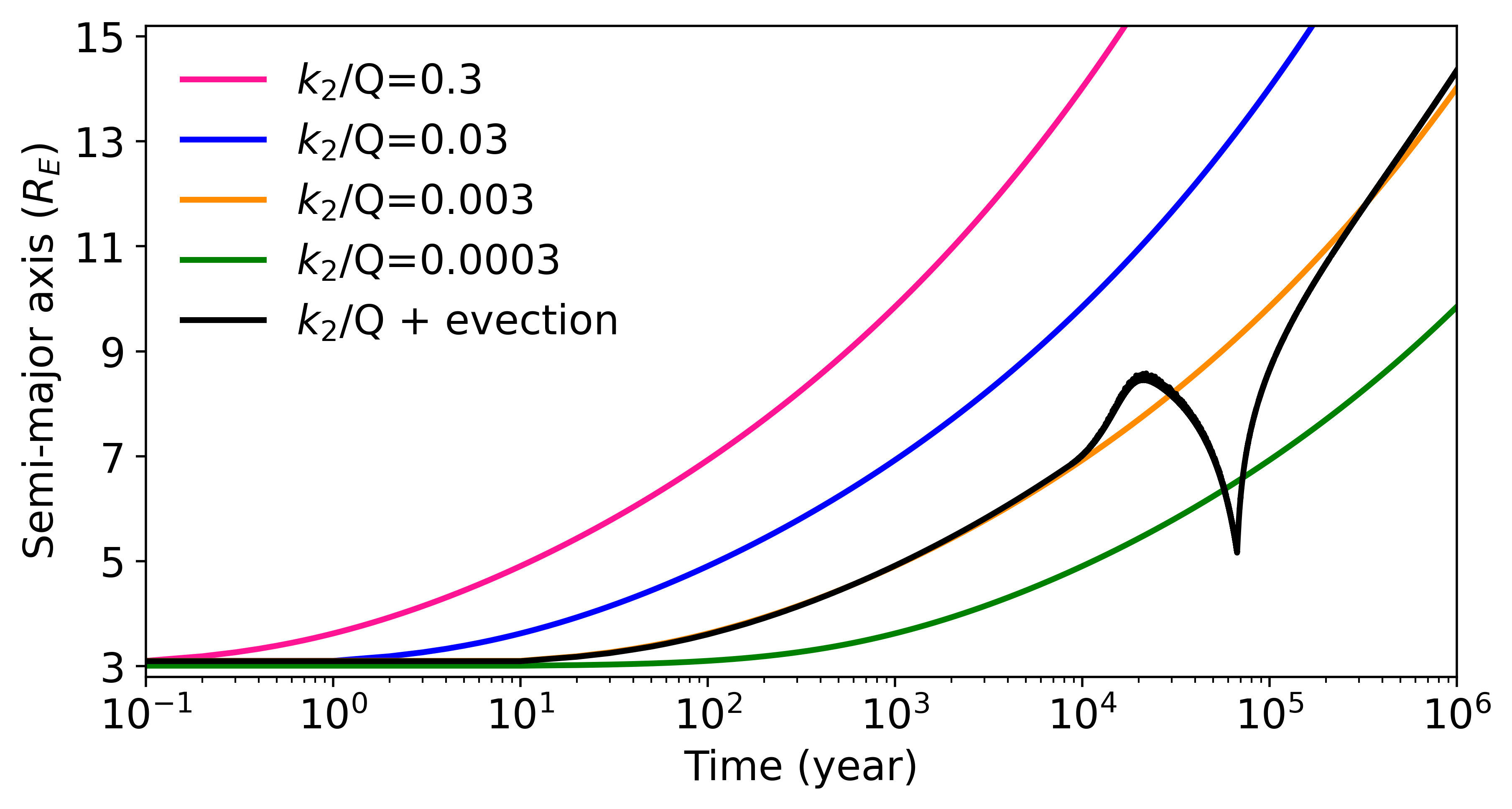}
\caption{Time evolution of the Earth-Moon distance. In the case of the colored lines, the Moon is migrating only due to the Earth's tides, with different values of $k_2/Q$. The black line, on the other hand, corresponds to the scenario in which the Moon is also involved in an evection resonance, as shown in Figure 3 of \cite{cuk2012making}.}
\label{fig:tides}
\end{figure}

In this section, we take into account the migration of the Moon due to tidal torques and calculate its composition over time. The dynamic evolution of the early Moon is a puzzle and various scenarios have been explored over the years \citep{webb1982tides,cuk2012making,cuk2016tidal,cuk2021tidal,tian2017coupled}. In particular, the uncertainty about the tidal effects on a molten Earth and Moon turns the post-formation evolution of the satellite into an intricate question \citep{chen2016tidal}. For simplicity, we adopt a simplified constant-Q tidal model for the evolution of the semi-major axis, described by \citep{goldreich1966q}:
\begin{equation}
\frac{da_M}{dt}=3\frac{k_2}{Q}\frac{M_M}{M_E}\sqrt{\frac{GM_E}{a_M}}\left(\frac{R_E}{a_M}\right)^5    
\end{equation}
where $k_2$ and $Q$ are the Love number and quality factor of Earth, respectively. Currently, these values are $k_2=0.3$ and $Q\approx10$ \citep[$k_2/Q\approx$0.03,][]{williams2015tides}. Whilst the quality factor of a molten Earth is unknown, it is believed to be higher at that time due to lower internal dissipation \citep{chen2016tidal,charnoz2021tidal}. Simultaneously, \(k_2\) is expected to have been larger due to the absence of a shear modulus in the planet \citep[e.g., see][]{raevskiy2015reconciling}. Seen this, we investigate the early Moon's evolution by assuming different values of the ratio \(k_2/Q\), specifically \(k_2/Q\)=0.3, 0.03, 0.003, and 0.0003. In addition, we explore the scenario proposed by \cite{cuk2012making}, in which the Moon is temporarily captured in evection resonance. Such a resonance occurs when the precession period of lunar perigee is equal to the orbital period of the Earth. For this scenario, we adopt the time evolution given in Figure 3 of \cite{cuk2012making}, for which $k_2/Q\approx 0.003$. The time evolution of the Moon's semi-major axis is shown in Figure~\ref{fig:tides}.

In the several tide migration scenarios considered, we account for the material lost from the Moon over time by linearly interpolating the net loss fluxes shown in Fig.~\ref{fig:magmaocean}. Our particular interest lies in the abundances of Na and K, which are evaluated based on the fractions of atoms shown in Table~\ref{tab:bse} \citep[see][]{charnoz2021tidal}. The analysis of the abundance of moderately volatile elements is interesting because they are measurable in limited amount of materials \citep{wolf1980moon}, allowing a direct comparison between our results and laboratory data.

\begin{figure}[]
\centering
\includegraphics[width=\columnwidth]{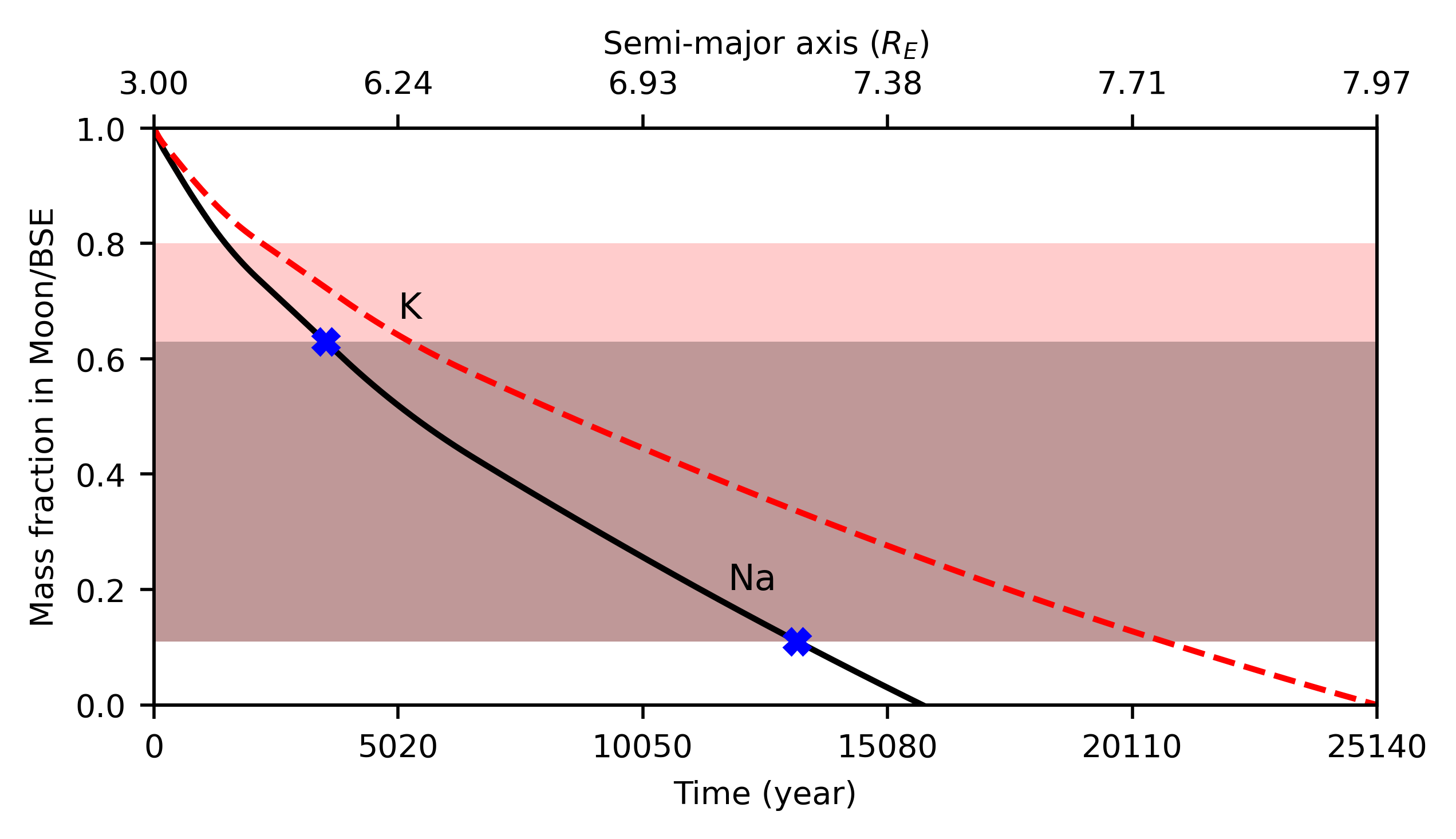}
\caption{Mass fractions of Na (black solid line) and K (red dashed line) in the Moon, in relation to BSE. The lunar surface temperature is 1800~K and the Earth's tidal parameter is 0.003. The time evolution is given on the lower x-axis, while the respective semi-major axis of the Moon is shown on the upper x-axis. The black and red regions outline, respectively, the ranges of Na and K mass fractions estimated in \cite{visscher2013chemistry}. The markers delimit the interval of time and semi-major axis in which the mass fraction of both elements simultaneously match the estimates. \label{fig:fNaK}}
\end{figure}

In Figure~\ref{fig:fNaK}, we present the fractions of Na and K in relation to BSE as a function of time, for a scenario in which $T_M$=1800~K and $k_2/Q$=0.003. Initially, the Moon's composition reflects that of the BSE and the mass fractions are 1. Over time, the Moon migrates outwards and loses material through hydrodynamic escape. The abundance of sodium, initially greater than that of potassium, decreases more rapidly. \cite{charnoz2021tidal} obtains that the loss of volatile via hydrodynamic escape ceases before the Moon reaches a distance of 6 $R_E$, leaving a content of moderately volatile elements in the Moon potentially similar to the estimates. According to \cite{visscher2013chemistry}, the current abundances of Na and K relative to BSE in the Moon vary within the ranges of 0.11-0.63 and 0.11-0.8, respectively.

\begin{figure}[]
\centering
\subfloat[]{\includegraphics[width=\columnwidth]{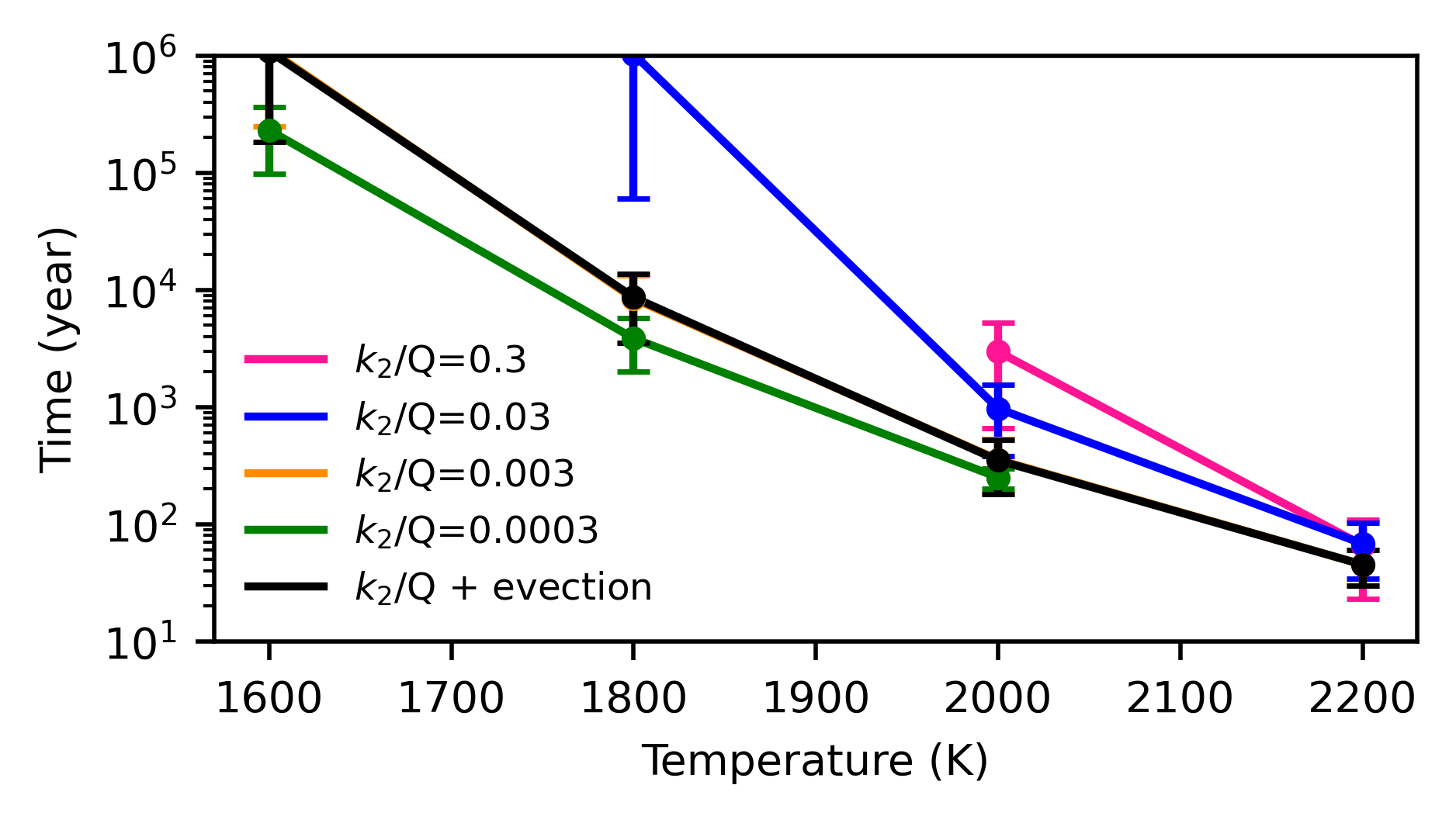}}\\
\subfloat[]{\includegraphics[width=\columnwidth]{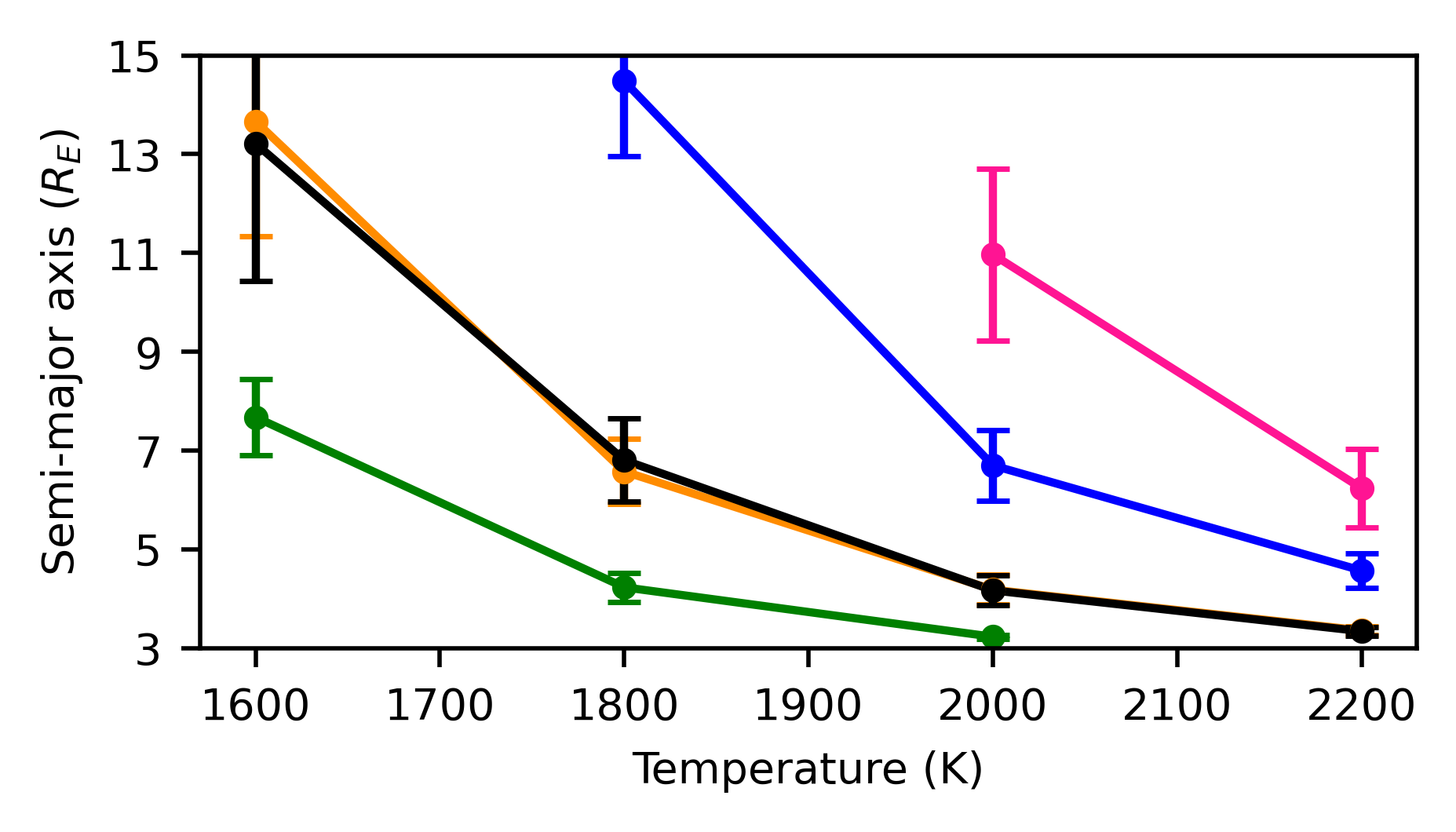}}\\
\caption{(a) Time interval for the end of volatile loss and respective (b) lunar positions for our entire set of simulations, with the Moon starting at 3 $R_E$. The x-axis shows the temperature of the lunar surface, while the different colors correspond to different Earth tidal parameters. Our calculations were carried out up to $10^6$ years.  \label{fig:solidification}}
\end{figure}

In our analysis, the net loss flux decreases more slowly with distance than observed in \cite{charnoz2021tidal} and we obtain in the simulation shown in Fig.~\ref{fig:fNaK}, the Moon being totally depleted of Na and K in less than 25,000 years. This suggests that mechanisms other than lunar migration must be at work in the system to prevent the total loss of volatiles. As will be discussed in Section~\ref{sec:lid}, the formation of a stagnant lid could be one of these mechanisms. Figure~\ref{fig:solidification} gives, for each of our simulations, the time interval and the corresponding lunar positions at which the loss of volatile must be stopped, in order to explain the observed lunar abundances of Na and K.

This time interval corresponds to the period during which the mass fractions of Na and K simultaneously match the estimates of \cite{visscher2013chemistry}. The midpoint of this range is calculated as the average between the minimum and maximum values. Our analysis extends up to $10^6$ years. As expected, higher temperatures on the Moon's surface result in faster volatile loss, requiring an earlier halt to this process. At the same time, lower $k_2/Q$ leads to slower migration, which means that the satellite spends a longer period closer to Earth. Therefore, in scenarios with lower $k_2/Q$, the interruption of volatile loss also needs to occur earlier. For $k_2/Q$=0.003, the time interval for the loss of volatile remains similar, regardless of whether the Moon is in evection resonance or not. This phenomenon occurs because the Moon loses most of its volatile before being captured in the resonance. For a surface temperature of 1400 K, it is observed that the Moon never loses enough of its volatile content. 

\begin{figure}[]
\centering
\includegraphics[width=\columnwidth]{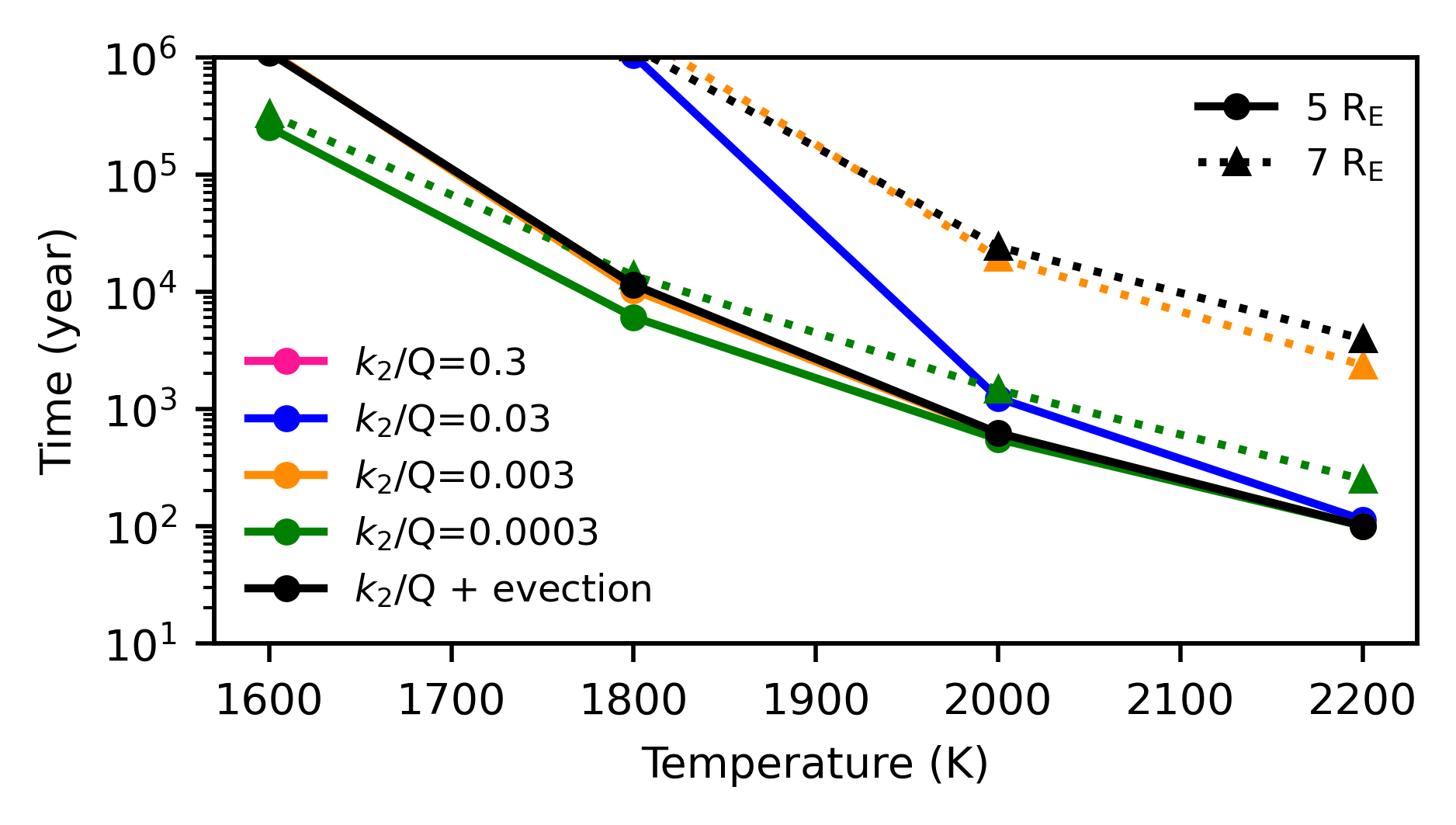}\\
\caption{Average time for the end of volatile loss for cases where the Moon starts at 5 $R_E$ (solid lines with circles) and 7 $R_E$ (dotted lines with triangles). The lunar surface temperature is shown on the x-axis, and the different colors represent different Earth tidal parameters. \label{fig:solidification_com}}
\end{figure}

In the scenario of the Moon forming from a proto-lunar disk, although the first aggregates coalesce at about 3~$R_E$, the Moon fully forms only at around 5~$R_E$ \citep{salmon2012lunar}. Seen this, we performed the same analysis as in Figure~\ref{fig:solidification}, but starting with the Moon at 5~$R_E$. To account for the direct formation scenario, we also considered a case where the Moon begins at 7~$R_E$, a location consistent with those obtained in \cite{kegerreis2022immediate,meier2024systematic}. The average times for the cessation of volatile loss in these two cases are shown in Figure~\ref{fig:solidification_com}. For the Moon at 5~$R_E$, the cessation times shows only a twofold increase, maintaining the same order of magnitude as those represented in Figure~\ref{fig:solidification}. For a Moon starting at 7~$R_E$, the volatile loss flux is significantly lower, and in cases of rapid tidal migration, the Moon does not lose sufficient volatile material. The Na and K mass fractions match the estimates only for $k_2/Q \leq 0.003$.

However, we draw the reader's attention to the fact that these results apply to a Moon in a circular orbit. For the disk formation scenario, this approximation is reasonable, as the Moon is expected to form with $e\sim 0.01$ \citep{salmon2012lunar}. For the direct formation scenario, in contrast, this assumption may not hold, as the Moon is expected to form in a highly eccentric orbit \citep[$e \sim 0.3$,][]{meier2024systematic}. Assuming a semi-major axis of 7~$R_E$ and $e = 0.3$, we performed a simple estimation of the mass loss over one orbital period by interpolating the flux at different orbital locations. This analysis indicates an approximately 30\% increase in mass loss compared to the circular case, suggesting that the cessation time of a directly formed Moon might be shorter than that shown in Figure~\ref{fig:solidification_com}, although we do not expect the values to differ by orders of magnitude. Nonetheless, this should be further investigated in future publications. In the next section, we calculate the timescales for the crystallization of the stagnant lid.

\section{Formation of an early stagnant lid on the Moon} \label{sec:lid}

The formation of a conductive lid over a vigorously convecting magma ocean on the Moon may result both from the buoyancy of suspended crystals within the magma during mantle solidification \citep{elkins-tanton11,elkins-tanton12} or the establishment of a stagnant lid due to rapid surface cooling into a magma with strongly temperature-dependent viscosity \citep[e.g.][]{davaillejaupart93}. This conductive lid can effectively extend magma ocean cooling duration \citep[e.g.][]{watsonetal22,michautneufeld22,lichtenbergetal23} and has the potential to reduce or even prevent volatile degassing.

\citet{elkins-tanton11} estimate that the formation of an anorthite flotation crust (plagioclase) on the lunar magma ocean should happen on a timescale of $\sim10^3$ years, assuming instantaneous crystal flotation. Anorthite floats due to its lower density compared to the interstitial liquid. Nonetheless, anorthite's contribution is negligible for the formation of a conductive lid until solidification of the lunar magma ocean reaches $\sim$80-90\% or a depth of $\sim$100~km. A stagnant lid that emerges early enough could limit heat loss from the magma ocean and therefore delay the formation of the anorthite crust \citep{pereraeal18,michautneufeld22}. Here, to estimate the onset of the conductive lid, we calculate the thermal evolution of the lunar magma ocean alongside the growth of a stagnant lid. Details about our model for stagnant lid formation are provided in \ref{sec:modellid}.

By integrating Equation~\ref{eq:heatflux}, we obtain the thermal evolution of the lunar magma ocean over $10^5$ years, assuming logarithmically spaced timesteps. At each interval, we assess the criteria expressed in Equation~\ref{eq:lidcond} for stagnant lid formation. Upon matching this criterion, a stagnant lid forms and the heat flux extracted from the magma ocean due to convection decreases. Our preliminary calculations show that the criterion is reached quite early, around $10$~years, when the magma ocean has a potential temperature of $\approx 2800$~K. However, the thickness of the stagnant lid developed during such a few years is on the order of $10^{-7}$~m, which in practical terms is negligible. 

To address this limitation, we propose that the stagnant lid becomes effective in inhibiting degassing and reducing magma ocean convection with the surface once it reaches a certain size, $h_0$. Determining this critical size falls beyond the scope of this paper; hence, we consider $h_0$ to span three orders of magnitude from $1$~m to $100$~m. Therefore, for every timestep in which Equation~\ref{eq:lidcond} holds true, we calculate the size of the stagnant lid $h$ via Equation~\ref{eq:lidsize}. Also, when $h \geq h_0$, we alter the convective heat flux in the magma ocean from $F(\Delta T)$ to $F(T_{\mu})$.

\begin{figure}[]
\centering
\includegraphics[width=\columnwidth]{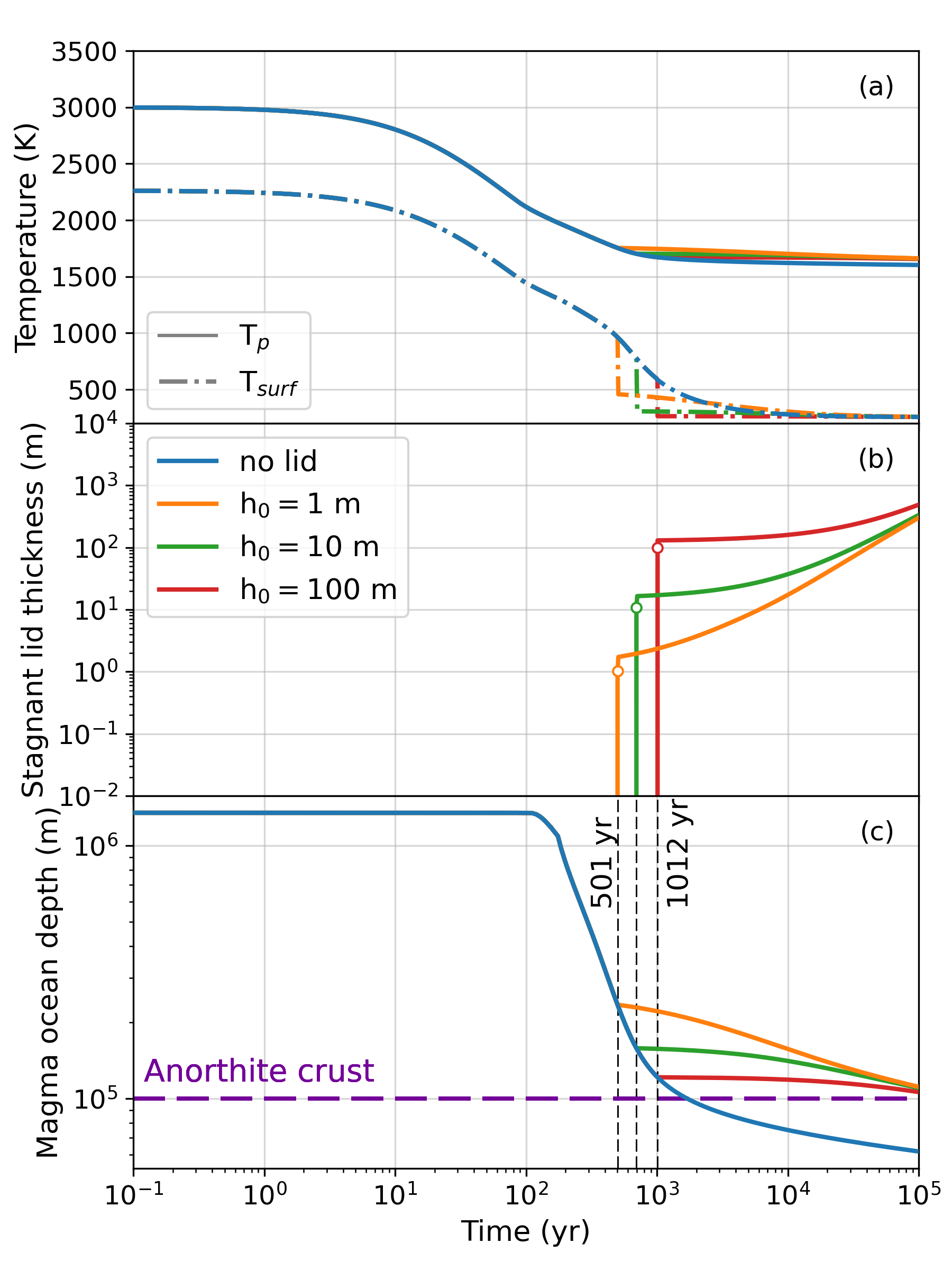}
\caption{Temporal evolution of (a) temperatures, (b) stagnant lid thickness, and (c) depth of the lunar magma ocean. The color-coding represents numerical calculations with lid emergence of sizes of 1~m, 10~m, 100~m, and without a lid. The purple dashed line denotes the start of anorthite solidification when magma ocean depth drops below 100~km. \label{fig:lids}}
\end{figure}

Figure~\ref{fig:lids} illustrates the evolution of the lunar magma ocean with an initial potential temperature of 3000~K for a case without stagnant lid formation and for three different values of $h_0$. The blue lines in Figure \ref{fig:lids}(a) denote the evolution of potential temperature (solid line) and surface temperature (dash-dotted line) for the case with no lid formation, in which the temperatures show an evolution similar to those obtained for planets without an atmosphere \citep[e.g.][]{lebrunetal13,nikolaouetal19}. Figure~\ref{fig:lids}(c) displays the evolution of magma ocean depth. To calculate the depth, we first obtain the radial temperature profile $T(r, t)$ given by the adiabat with $T_p(t)$, which also provides the solid fraction in the magma ocean. Then, we assume that all formed solids sink and instantly settle in the liquid magma, adding to the magma ocean's bottom. We can estimate the onset of anorthite crust formation when the magma ocean depth reaches 100~km \citep[e.g.][]{elkins-tanton11,tian2017coupled}. In Figure~\ref{fig:lids}(c), the purple dashed line indicates the onset of anorthite crust according to this criterion.

We observe that the anorthite crust formation occurs in approximately 1800~years for the magma ocean evolving without stagnant lid formation, a finding consistent with multi-composition fractional crystallization models \citep{elkins-tanton11}. For $h_0=1$~m, lid emergence occurs around 501~years. At the time of lid emergence, Figure~\ref{fig:lids}(a) shows the surface temperature drops due to the decrease in convective heat flux from $F(\Delta T)$ to $F(T_{\mu})$ (see \ref{sec:modellid} for details) and the magma ocean's potential temperature remains slightly higher than in the case with no lid. Lid emergence for $h_0$ equal to 10~m and 100~m occurs at 698~years and 1012~years, respectively. In Figure~\ref{fig:lids}(c), we note that the onset of the stagnant lid precedes the beginning of anorthite solidification for all $h_0$ values considered. Yet, the early emergence of the stagnant lid delays plagioclase formation to the order of $10^5$~years.

For the sake of completeness, we have computed the case with $h_0 = 1$~km, obtaining that the time to form the lid is roughly the same as the time to form the flotation crust. For different initial potential temperatures, we find that the variation in the timescale for the emergence of the lid is marginal. This is explained by the fact that convection, and therefore the heat flux escaping from the magma ocean, is proportional to the mantle's potential temperature.

We also compare the convective timescales of the magma ocean with the timescales during which gas parcels remain cycling in the lunar extended atmosphere, which primarily occurs for $a_M\geq$3.2~$R_M$. The convective velocities of the magma ocean during most of the convecting-only phase is on the order of 1~m/s, resulting in overturn timescales of about 20 days. We calculate the timescale for a gas parcel to be re-accreted by Moon to be about 0.4 day, depending on the assumed lunar surface temperature. This implies that $\sim$50 cycles of gas in atmosphere occur before the magma ocean undergoes a complete overturn and renews its surface.

Given the simplicity of the model used, the strong conclusion of this section is that, regardless of the stability of the stagnant lid over the lunar magma ocean, the different scenarios indicate that volatile loss should be inhibited in a timescale of $10^3$~years. This timescale is consistent with our analysis of the end of volatile loss (Figure~\ref{fig:solidification}), provided that the lunar temperature is around 1800~K-2000~K. This result is particularly valid for \(k_2/Q\) values lower than that of the present-day Moon, while for equal or higher values, the best match occurs at temperatures around 2000~K. For the direct formation scenario, in which the Moon forms at 7~$R_E$, $k_2/Q\approx$0.0003 is required to explain the concentrations of Na and K in the satellite, with the best matches occurring for temperatures around 2000~K.

It should be noted that the various temperatures employed are derived from distinct problems and assumptions. The lunar temperature of previous sections is a mean temperature of the magma ocean and is fixed, while both potential and surface temperatures in this section vary in time. Nonetheless, by comparing the time-weighted average potential temperature of the magma ocean ($\bar{T_p}$), as discussed herein, with the lunar temperature ranging from 1800~K to 2000~K, we find good correspondence. For the smallest $h_0$, $\bar{T_p}$ is equal to 1990.77~K and when $h_0$ is set to 100~m, $\bar{T_p}$ gives 1844.11~K.

\section{Discussion} \label{sec:discussion}
In this paper, we put to trial the hypothesis of the tidally-driven atmospheric escape scenario. \cite{charnoz2021tidal} assume a 1D steady-state approach and apply the hydrodynamic escape formalism to the Earth-Moon line to calculate the material crossing the lunar Hill sphere. Here, we evaluate the problem in a more general way by performing 2D time-dependent hydrodynamic simulations, which allows us to track the dynamic evolution of the volatile gas. As predicted by \cite{charnoz2021tidal}, we find that when the Moon is closer to the Earth, most of the gas material that crosses the Moon's Roche lobe does not return to the satellite. The fraction of re-accretion by the Moon is less than 30\% when the satellite is located up to 3.5 $R_E$. Our results therefore contradict those of \cite{dauphas2022extent}, which state that hydrodynamic escape from the Moon is insufficient to explain the depletion of moderate volatile elements on the satellite. However, the physics we consider is more realistic as we discuss below. 

\subsection{On the work of \cite{dauphas2022extent}}

\cite{dauphas2022extent} reported a critical comparison of different models of volatile loss. By using the Hertz-Knudsen (HK) evaporation theory -- which describes the kinetic evaporation of a gas from a free surface --, they link the measured isotopic fractionation to the evaporating flux \citep[see, e.g.][for a detailed model]{Richter_2002}. In fact, the evaporation flux ($J_i$) in the kinetic theory of gases is proportional to the difference between the gas pressure (P) and the equilibrium saturating vapor pressure of the liquid ($P_{sat}$), so for a species i, $J_i=\gamma_s(P-P_{sat})/(2\pi\mu kT)$, where $\gamma_s$ is a sticking efficiency coefficient -- measured experimentally and between 0 and 1 -- and $\mu$ is the mass of the molecule \citep[see, e.g.][]{hirth1963condensation,Richter_2002}. Different isotopes with slightly different values of $\mu$ will have slightly different evaporating fluxes and different vapor pressures. These different evaporation fluxes induce a net isotopic fractionation of the material that remains in the liquid, leading to progressive enrichment in the heavier isotopes. Based on this argument, \cite{dauphas2022extent} proposed that the measured isotopic variations of K in lunar rocks imply that the effective vapor pressure above the lunar magma must be $P\sim0.99P_{sat}$ (with $T\sim1700$~K). Due to the close-to-total saturation and low temperature, it is concluded that about 10 Myrs would be necessary to explain the measured isotopic composition, a time by which a stagnant lid would have formed ($\sim$1000 years).

Whereas we do not dispute the simple calculations presented in \cite{dauphas2022extent}, the approach developed in this paper and in \cite{charnoz2021tidal} relies on completely different physics: the Navier-Stokes equation -- conservation of mass, energy, and momentum -- is solved and the net flux at the Hill sphere of the Moon is computed, not simply at the magma ocean surface as in the HK model. Many additional physical complexities arise that make the HK calculation too simplistic: for example, the escaping flux is highly anisotropic, potentially leading to non-isotropic fractionation and mixing in the lunar atmosphere before escape. Also, molecular and turbulent isotopic diffusion in the lunar magma ocean and the atmosphere could alter isotopic fractionation before escape (see Figure \ref{fig:angmajor}). Finally, Hertz-Knudsen equation is derived from the kinetic theory of gases and is adapted to evaporation from a free liquid surface with a vapor with zero net velocity \citep{hirth1963condensation}. To account for overly simplistic physics, the HK theory uses several multiplicative factors to correct for  physics not included in the equation (e.g. the sticking coefficient $\gamma_s$). In the problem considered in \cite{charnoz2021tidal}, the gas flow is forced by a strong pressure gradient -- this is a Parker Wind -- with non-zero net velocity at the magma ocean surface, which is a priori outside the domain of validity of the HK theory. Therefore, the kinetic fractionation of HK developed in the \cite{dauphas2022extent} model does not capture all the complexity of the escape process presented in \cite{charnoz2021tidal}.

This timescale discrepancy between dynamical processes and atomic fractionation processes is not limited to the tidally assisted escape model. In \cite{dauphas2022extent}, it is shown that viscous drainage from a disk, using the MRI model of \cite{charnoz2015evolution}, could lead to an isotopic fractionation of K compatible with lunar samples in less than 400 days. However, dynamical studies of lunar disk cooling and spreading show that spreading and cooling timescales are about 50-100 years \citep[see][]{Thompson_Stevenson_88, Ward_2012_vert_structure}, leading again to an apparent paradox where dynamical and isotopic fractionation timescales are very different. 

In addition, the viscous drainage model requires an ionized environment \citep{charnoz2021tidal} with a non-zero local magnetic field to foster the magnetorotational turbulence. Isotopic fractionation processes in an ionized environment are not well understood and can produce mass-independent fractionation processes \citep{Kuga_2017, Robert_2021}, which can further complicate the problem. We conclude that it is still very difficult to reconcile the fractionation timescale deduced from the mere application of the HK theory with dynamical models of hydrodynamical escape or lunar disk evolution. Improvements could be made on both sides. Thus, for now, this question remains open.

\subsection{Comparison with \cite{charnoz2021tidal} and General Remarks}

In spite of the differences in methodology and in the properties of the lunar magma ocean between our work and that of \cite{charnoz2021tidal}, we obtain escape fluxes that are fairly consistent. This is especially evident, at least in order of magnitude, with those reported by \cite{charnoz2021tidal} when the Moon is at about 3~$R_E$. The Earth's tides reduce the energy needed for thermal escape and, closer to the Earth, it turns out that all the gas content formed by the evaporation of the magma ocean is energetic enough to cross Moon's Roche lobe and be placed in orbit around the Earth. Only a fraction of this gas is re-accreted by the Moon, resulting in an efficient net loss of volatiles. 

As the Moon migrates outwards due to tidal torques with the Earth, the energy required for thermal escape increases and less material is able to cross Moon's Roche lobe. However, we find a much smoother drop in the loss flux with distance than \cite{charnoz2021tidal}. While the authors conclude that the loss of volatiles from the Moon is halted up to 6~$R_E$, we obtain that, at this position, the net loss flux is reduced by only one order of magnitude (Fig. ~\ref{fig:magmaocean}). This implies that tidally-driven atmospheric escape is even more efficient than previously estimated. Regardless of whether the loss of volatiles starts already in the lunar building blocks, in which the loss flux should be even more efficient due to their smaller masses, or starts only with the fully formed Moon at around 5~$R_E$, the loss is expected to be strong enough to completely deplete the objects in volatiles in Na and K, if the surface temperature is above 1600~K.

In an isothermal expanding gas cloud, a fraction of the volatiles is expected to condense due to pressure changes as the gas flows toward the lunar Roche lobe, an effect not considered in this study. Nonetheless, condensate droplets small enough ($\lesssim \mu$m) are expected to remain well-coupled to the gas, being carried outward by inertia. Furthermore, if the early Earth's photosphere had temperatures $\gtrsim 2000$~K, micrometer-sized condensates would have been subjected to radiation pressure, possibly being blown out of the system \citep{hyodo2018impact}. Therefore, we expect that condensation would increase the loss flux, making hydrodynamic escape even more efficient. Indeed, \cite{charnoz2021tidal} showed that this effect generates a 40\% increase in loss flux, and we leave the question of how condensation quantitatively impacts the net loss flux for future publications.

Additionally, Earth tides would distort the Moon, giving it a non-spherical shape \citep{chandrasekhar1989equilibrium}. For example, assuming the satellite is at 3~$R_E$ and approximating it as a perfect fluid, we calculate that its projection in the XY plane corresponds to an ellipse with semi-axes of 0.95 and 1.15 Moon radii \citep[see][]{madeira2023exploring}. In general, planetary tides increase the satellite's surface area, which leads to larger ejection fluxes. Test simulations accounting for the Moon's distortion show that the net volatile flux can increase by approximately 30\% at 3~$R_E$. Therefore, Moon's distortion is also expected to make tidally assisted hydrodynamic escape more efficient, even though it is not expected to change the order of magnitude of our results.

These results highlight the need for another mechanism, in addition to the Moon's orbital evolution, to explain the abundances of moderate volatile elements measured in lunar mare basalts. We suggest that a conductive lid could develop over the magma ocean, stopping the loss of volatiles and surface recycling, either due to a high temperature contrast -- thermal stagnant lid -- or the accumulation of buoyant solids -- anorthite flotation crust. Considering a simple model of thermal evolution and stagnant lid growth, we find that a stagnant lid ranging in size from 1~m to 100~m should form on top of the convective layer within an interval of about 500-1000 years. If this thermal stagnant lid is stable, it would block magma ocean degassing within 1000~years. However, if the stagnant lid is not stable, the lunar magma ocean is expected to lose heat more rapidly, initiating anorthite solidification and the formation of a flotation crust around 1800~years. Regardless of the scenario envisioned in our model, volatile loss should stop on a timescale of $10^3$~years. To achieve such timing, we constrain the temperature of the early Moon's magma ocean between 1800-2000~K., which is in line with models based on the Cr isotopic composition of the Moon \citep{sossi2018volatile}.

For the lunar positions $\gtrsim$3.2~$R_E$, we obtain that the material is predominantly re-accreted on the trailing side of the Moon, which, in principle, could result in a dichotomy of volatile element concentrations between the east and west sides of the satellite. This is only valid if the satellite lost most of its volatile content when fully formed, since the impacts between its building blocks would erase any possible dichotomy they might have had. In addition, other mechanisms might act over time to reduce this dichotomy, such as the redistribution of surface material due to meteorite impacts and the reorientation of lunar longitudes due to migration or other events -- the Moon's current zero longitude meridian does not necessarily correspond to the same one from billions of years ago. 

However, if measured, the dichotomy in the content of volatiles in the Moon could give clues about the migration of the satellite. For the same temperature, the time needed to obtain the current abundances of Na and K in the satellite are of the same order, regardless of tidal migration (Fig~\ref{fig:solidification}). However, the location of the Moon at this time depends on the migration parameter. For the Moon's current $k_2/Q$ and temperature of 2000~K, gas loss is predominantly away from the Earth, and more than 99\% of the material is re-accreted on the trailing side of the Moon, implying a strong dichotomy. Now, for a slow migration with $k_2/Q=0.0003$, for example, the loss of volatiles occurs within 3.2~$R_E$, and only $\sim$60\% of the material is re-accreted on the trailing side.

\subsection{Caveats}
The early Moon's degassing is an intricate problem, with many potentially relevant effects being not accounted for in our model. These effects, however, should be explored in future studies. The first of these, discussed in the previous section, concerns the possible condensation of volatiles -- an effect that cannot be currently computed given that the \texttt{FARGOCA} code, in its present version, is not a multi-phase code. Also mentioned, the Moon distortion by Earth tides is another of these effects. We assume the Moon to be spherical in our simulations, which is a reasonable approximation only if the early Moon had a high internal cohesion \citep[see][]{madeira2023exploring}. However, the lack of knowledge of physical parameters of the early Moon and early Earth force us to always rely on assumptions regarding tidal distortion, and investigating the Moon's shape effect in the hydrodynamic escape process is beyond the scope of this work.

Another limitation of our model is related to the 2D approximation. Due to the high temperature of the material ejected from the Moon, a thick circum-Earth disk forms, similar to the expected to the proto-lunar disk \citep{nakajima2014investigation,ward2011vertical}. At the end of our simulations, the disk height at 3~$R_E$ reaches about 0.5 Moon radii, and at 8~$R_E$, it exceeds 2 Moon radii. While our 2D approximation may be marginally appropriate at 3~$R_E$, where hydrodynamic escape is strongest, we acknowledge that we may be overlooking some physical effects when the Moon is at greater distances from the Earth. The extended atmosphere of a Moon embedded in a thick disk is expected to exchange material with the circum-Earth disk not only through spiral arms (a 2D effect), but also via meridional gas circulation near the gap \citep[see details in][]{teague2019meridional,schulik2020structure}. These meridional flows deliver material to the satellite and consequently, the re-accretion flux onto the Moon might be higher than obtained by us. A more realistic model of the disk would require highly computationally expensive 3D hydrodynamical simulations, and our computations were already very demanding with the current setup.

We also mention the uncertainty surrounding the initial composition of the Moon. While it is well-established that the Moon and BSE formed with the same concentration of refractory elements \citep{ringwood1992volatile}, our assumption that the Moon has a BSE-like concentration of volatile elements is more speculative. To better understand how this assumption might impact the system, we recomputed the vapor composition by varying the mass fractions of Na and K in the magma ocean. Our findings show that the lunar surface density -- which determines the ejection flux -- is proportional to the concentration of these elements. For instance, when the mass fractions of Na and K are doubled, the lunar surface density increases by a factor of 1.4.

The lunar surface density impacts the ejection flux (Figure~\ref{fig:magmaoceana}), but it is not expected to significantly affect the fraction of material that is re-accreted by the Moon (Figure~\ref{fig:magmaoceanb}), implying that the net loss flux will also be proportional to the initial volatile content. Therefore, it is possible that the overall timescale for the end of volatile loss we obtained, at least in terms of order of magnitude, remains unchanged for different volatile contents in the magma ocean. This is because any initial excess or deficit of volatiles relative to the BSE might be balanced by corresponding changes in depletion rates. We will leave the exploration of this scenario for a future publication.

To conclude, we address a comment on the gas adiabatic index. The \texttt{VAPOROCK} code provides the actual adiabatic index of the vapour gas, which may differ from the value expected in the vertically integrated (2D) scenario simulated here. This difference arises because, in the latter, there is an additional degree of freedom in the system due to vertical disk expansion \citep{goldreich1986stability}, and the 2D adiabatic index should be lower than the actual value used in our simulations. To quantify the impact of this parameter, we calculated the net loss flux for $T_M = 1800~\mathrm{K}$ and $a_M = 3$~$R_E$, varying the adiabatic index. A 10\% reduction in this parameter increases the flux by an order of magnitude, whereas reducing the flux by the same amount requires a 60\% increase in the adiabatic index. Within the range $0.9\gamma - 1.6\gamma$ (1.105–1.964 for $T_M = 1800~\mathrm{K}$ and $a_M = 3$~$R_E$), the volatile loss timescale is expected to remain of the same order of magnitude as the results presented here, leaving our conclusions unaffected.

\section{Conclusion} \label{sec:conclusion}

We conclude that tidally driven atmospheric escape is a solid mechanism to explain the depletion of moderate volatile elements in the Moon, as it does not rely on any specific formation model. Our model is able to explain the current observed depletion of Na and K on the satellite, without requiring any prior loss of volatiles in the proto-Moon disk, which is uncertain to have existed \citep{kegerreis2022immediate,Sossi2024}. Our results show that the hydrodynamic escape of the early Moon is more intense than previously predicted by \cite{tang2020evaporation,charnoz2021tidal,dauphas2022extent}, with the formation of a stagnant lid being the key mechanism preventing the complete depletion of Na and K in the satellite. To reproduce the current abundances of these elements, the temperature of the Moon's magma ocean must be around 1800-2000~K.

\section*{Acknowledgments}
GM and FM thanks the European Research Council (\#101001282, METAL). LE acknowledges the financial support from FAPESP through grants 2021/00628-6 and 2023/09307-3. HPC resources from “Mesocentre SIGAMM”, hosted by Observatoire de la Côte d’Azur were used. We thank Alain Miniussi for maintaining and developing the code FARGOCA and Alessandro Morbidelli, Aurelien Crida, Angela Limare, and Maylis Landeau for helpful discussions. We thank the anonymous reviewers for their comments, which greatly helped us to improve the article.

\bibliographystyle{elsarticle-harv} 
\bibliography{article.bib}


\appendix
\section{Magma ocean model} \label{sec:modellid}

\begin{figure}[]
\centering
\includegraphics[width=\columnwidth]{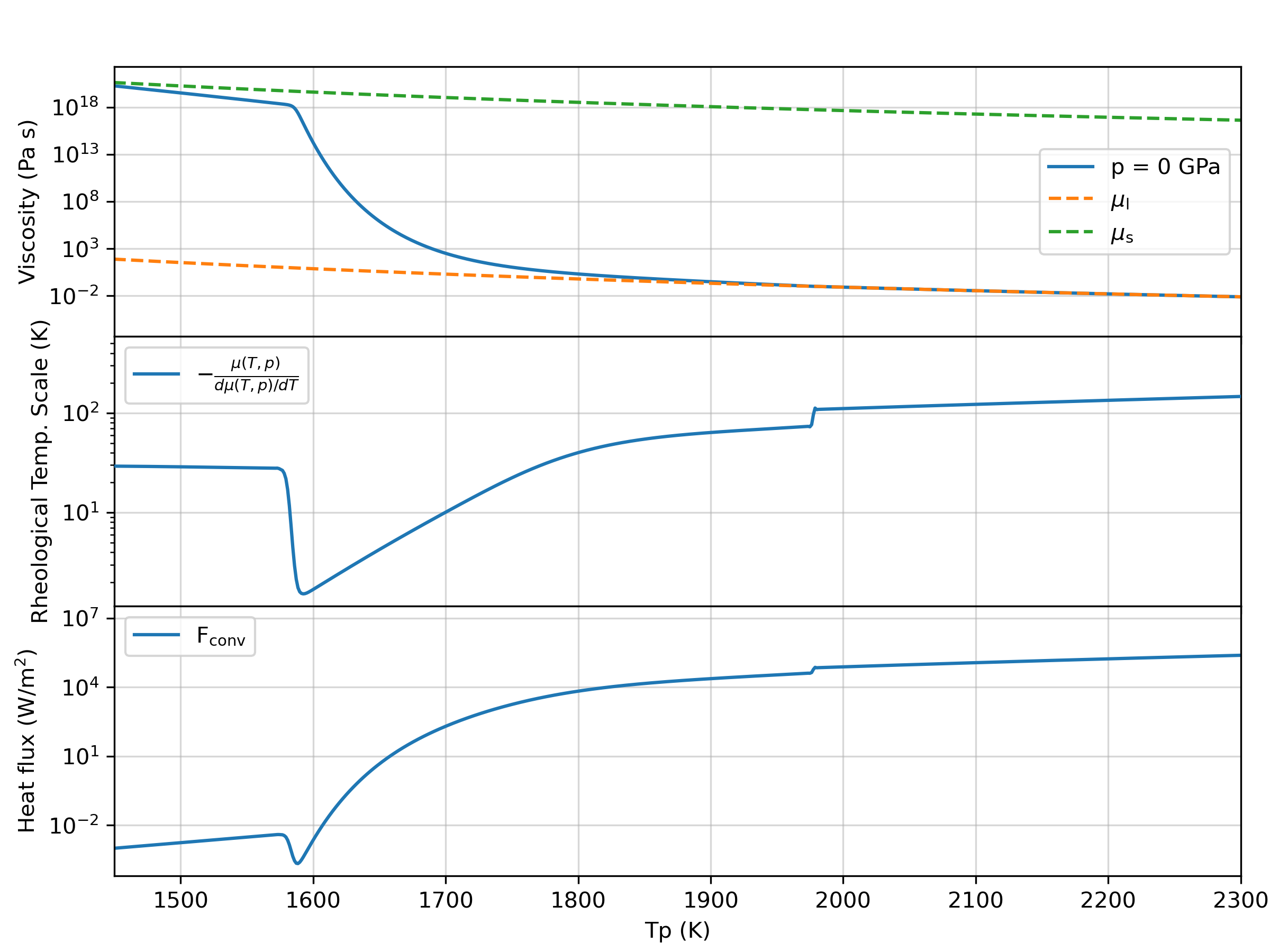}
\caption{Viscosity, rheological temperature scale and convective heat flux in function of the potential temperature. \label{fig:visc}}
\end{figure}

In this work, we adapted the model of mixed interior convection below a growing conductive stagnant lid from \citet{watsonetal22} to start from a convecting-only lunar magma ocean without a lid. Table~\ref{tab:lid} describes the general parameters used in the model. As in \citet{watsonetal22}, we adopt solidus and liquidus temperatures linearized with respect to pressure $p$ \citep{maximeeal17}, while the viscosity for the liquid and solid phases of the magma follows the Arrhenius law, where the viscosity contrast between solid and liquid is on the order of $10^{22}$~Pa$\cdot$s. In the mixing phase, the viscosity remains continuous following the function $f(\phi)$ of \citet{costaetal09}. The top panel of Figure~\ref{fig:visc} shows the viscosity function for $p = 0$~GPa. Empirical constants $\delta$, $\gamma$, and the critical solid fraction $\phi^*$ are chosen as in Table~\ref{tab:lid}, and $\xi$ is used as a free parameter to ensure viscosity continuity at the limit $\phi \to 1$.

\begin{table*}[]
\centering
\caption{Stagnant lid evolution model quantities. \label{tab:lid}}
\begin{tabular}{lccl}
\hline\hline
  Symbol   & Value & Description & Reference\\ \hline\\
\multicolumn{4}{c}{MOON PARAMETERS} \\ \hline
$R$  &   1737    &   Moon radius (km) &  \multirow{4}{*}{\citet{garciaetal11}}  \\
$R_{c}$  &   380    &  Moon core radius (km) &   \\
$\rho$   &  3328     &  mantle mean density (kg~m$^{-3}$) &   \\
$\rho_c$   &  5171     &  core mean density (kg~m$^{-3}$) &   \\
$g(r)$  &   $\frac{G[M_{\rm{core}} + M_{\rm{mantle}}(r)]}{r^2}$  &  Moon gravity a radius $r$ (m~s$^{-1}$) & \\\\
\multicolumn{4}{c}{VISCOSITY PARAMETERS} \\ \hline
$T_{\rm{liq}}$  &   $1400 + 149.5p$    &  \multirow{2}{*}{silicate liquidus and solidus. $p$ (GPa), $T$ (K)} & \multirow{5}{*}{\citet{maximeeal17}} \\
$T_{\rm{sol}}$  &   $1977 + 64.1p$    &   &            \\
$\mu_{\rm{l}}$  &   $10^{21} \exp\left [ \frac{E_{\mu}}{R_g} \left ( \frac{1}{T_p} - \frac{1}{T_{\rm{liq}}}\right )\right ]$   & solid phase viscosity (Pa~s) &  \\
$\mu_{\rm{s}}$  & $0.1 \exp\left [ \frac{E_{\mu}}{R_g} \left ( \frac{1}{T_p} - \frac{1}{T_{\rm{sol}}}\right )\right ]$ & liquid phase viscosity (Pa~s) &  \multirow{5}{*}{\parbox{3.1cm}{\citet{watsonetal22} \citet{costaetal09}}}\\
$\mu_{\rm{l+s}}$ &  $\mu_{\rm{l}} f(\phi)$   & partially molten phase viscosity (Pa~s) &  \\
$f(\phi)$       &   See \citet{costaetal09}   &   empirical function for relative viscosity. &   \\
$\phi*$ &  0.5  &  critical solid fraction &  \\
$\delta$, $\gamma$ &   7, 4    &  empirical constants for $f(\phi)$ &  \\
$B_E$   & 2.5 &  Einstein coefficient   &  \\\\
\multicolumn{4}{c}{MAGMA OCEAN PARAMETERS} \\ \hline
$\alpha_0$        &  $3 \times 10^{-5}$  &  thermal expansivity at $p=0$ (K$^{-1}$)  & \multirow{10}{*}{\parbox{3.3cm}{\citet{lebrunetal13} \citet{nikolaouetal19}}} \\
$\alpha$   &  $\alpha_0 (0.02 p + 1)^{-3/4}$   &  thermal expansivity (K$^{-1}$).  &  \\
$C_p$      &   $10^{3}$   &  specific heat capacity (K$^{-1}$) &  \\
$\kappa$        &   $8.8 \times 10^{-7}$    &  thermal diffusivity (m$^{2}$~s$^{-1}$) &  \\
$k$        &   $\kappa \rho C_p$    &  thermal conductivity (W~m$^{-1}$~K$^{-1}$) &  \\
$L$        &   $4 \times 10^{5}$   &  latent heat for silicate solidification (J~kg$^{-1}$) &  \\
$\epsilon$  &   1   & black body emissivity        &   \\
$a$  &   0.3   & Earth's albedo &  \\
$\sigma$  &   $5.670374\times 10^{-8}$  & Stefan-Boltzmann constant (W~m$^{-2}$~K$^{-4}$) &  \\
$S_0$  &   1361   & solar flux (W~m$^{-2}$)        &  \\
\hline
\end{tabular}
\end{table*}

Given that the magma ocean is convecting, the temperature profile follows an adiabat as in \citet{lebrunetal13}, but with a pressure range of 0 to $\sim$8 GPa for the Moon. The variation of temperature with respect to pressure is given by:
\begin{equation}
    \frac{dT}{dp} = \frac{\alpha T}{\rho C_p},
\end{equation}
where $T(p = 0)$ is the potential temperature $T_p$, $\alpha$ is the thermal expansivity, and $C_p$ is the specific heat capacity. By calculating the adiabatic temperature profiles for various $T_p$, it shows that even in the relatively low-pressure environment of the Moon's mantle, the adiabats first intersect the liquidus (and solidus) at the highest pressures. This indicates bottom-up solidification.

Integrating the energy conservation equation over the volume of the magma ocean ($V_m$), we obtain the evolution of the potential temperature in time \citep{abe97}:
\begin{equation}
    \int_{m} dV_{m} \rho C_p\frac{dT_p}{dt} = \int_{m} \left( -F + \rho L \frac{d\phi}{dt}\right)dV_{m}, \label{eq:thermalevo}
\end{equation}
with $L$ being the latent heat of solidification. The heat flux, $F$, follows the Rayleigh-Bénard convection equation and is given as a function of the temperature difference across the unstable boundary $\Delta T$ and the Rayleigh number $Ra$:
\begin{equation}
    F(\Delta T) = C \frac{k \Delta T}{d} Ra(\Delta T)^{1/3}, 
\end{equation}
where
\begin{equation}
Ra(\Delta T) =\frac{\rho \alpha g d^3}{\kappa \mu} \Delta T.   
\end{equation}
This leads to:
\begin{equation}
    F(\Delta T) = C k\left( \frac{ \rho \alpha g}{\kappa \mu} \right)^{1/3} \Delta T^{4/3}. \label{eq:heatflux}
\end{equation}
Here, $F(\Delta T)$ is independent of the height of the convective layer $d$. $k = \kappa \rho C_p$ is the thermal conductivity, $\kappa$ is the thermal diffusivity and $C$ is an experimental constant of proportionality. While the magma ocean evolves without a stagnant lid, the heat flux is measured immediately at the surface with $\Delta T = T_p - T_{\rm{surf}}$ representing the temperature difference at the top of the convective layer and $C$ assumes the value of 0.089 \citep{lebrunetal13}. Note that the dynamical viscosity $\mu$ is both temperature and pressure dependent due to our assumed formulation for viscosity, while $g$ and $\alpha$ can be considered as $g(r = 0)$ and $\alpha_0$, respectively.

We calculate the surface temperature by balancing the convective flux with the irradiating flux at the surface:
\begin{equation}
    F(\Delta T) = F_{\rm{rad}} = \sigma \epsilon (T_{\rm{surf}}^4 - T_{\rm{eq}}^4) \label{eq:boundarytsurf}
\end{equation}
where the equilibrium temperature $T_{\rm{eq}}$ is given in terms of albedo $a$, Stefan-Boltzmann constant $\sigma$ and Solar flux $S_0$ (see table \ref{tab:lid}):
\begin{equation}
    T_{\rm{eq}}^4 = \frac{(1 - a)}{4 \sigma}S_0.
\end{equation}
At each timestep, we compute the condition for the existence of the stagnant lid due to the viscosity contrast at the top of the convection layer, given by~\citet{davaillejaupart93}
\begin{equation}
T_{\mu} = -\frac{\mu(T,p)}{d\mu(T,p)/dT} < \frac{(T_p - T_{\rm{surf}})}{B}. \label{eq:lidcond}
\end{equation}

The above condition depends on the factor $B$ that can be constrained by laboratory experiments~\citep{davaillejaupart93} or numerical models~\citep{thirietetal19}. We use $B=2.54$ provided by~\citet{thirietetal19}. If the condition is satisfied, a stagnant lid forms at the top of the magma ocean reducing the heat flux from magma ocean convection and the surface temperature. In the model, the heat flux drop is represented by the fact that $F(\Delta T)$ is then calculated with $\Delta T = T_{\mu}$ which is smaller than $T_p - T_{\rm{surf}}$ and the constant $C$ becomes 0.47 as measured in the experiments of~\citet{davaillejaupart93}. Another consequence is a new boundary interface between the stagnant lid and the magma ocean at the depth of the stagnant lid $h$ as in~\citet{watsonetal22}. By assuming a temperature drop at the boundary between the magma ocean and the stagnant lid given by $T_{\mu}B$, we can obtain the relation for the temperature at the bottom of the stagnant lid, $T_{\rm{bsl}}$, as
\begin{equation}
T_{\mu}B = T_p - T_{\rm{bsl}}.
\end{equation}
The convective flux from the magma ocean $F(T_{\mu})$ is now balanced with the heat conduction flux through the stagnant lid $F_{\rm{cond}}$, and the growth of the stagnant lid is proportional to the difference $F_{\rm{cond}} - F(T_{\mu})$. The conductive flux can be calculated as

\begin{equation}
F_{\rm{cond}} = -k \eval{\frac{\partial T}{\partial z}}_{z~=~R - h}.
\end{equation}

As we are only interested in the time period during which the stagnant lid emerges and has  very small size relative to the convection layer, we consider that the timescale for heat diffusion through the lid is very short so that the temperature profile through the lid is approximately linear. This implies that
\begin{equation}
F_{\rm{cond}} = -k \frac{(T_{\rm{surf}} - T_{\rm{bsl}})}{h} \approx F(T_{\mu})
\end{equation}
Then, 
\begin{equation}
h = -k \frac{(T_{\rm{surf}} - T_{\rm{bsl}})}{F(T_{\mu})}.
\label{eq:lidsize}
\end{equation}
This approximation does not produce significant differences in the calculation of the stagnant lid thickness $h$, and it is much faster compared to integrating a heat conduction equation for such a thin stagnant lid ($< 1$~km). Note that due to this approximation, $F(T_{\mu})$ remains the heat flux reaching the surface and therefore is used to determine $T_{\rm{surf}}$.

In the Figure~\ref{fig:visc}, we plot the viscosity $\mu(T,p)$, rheological temperature scale $T_{\mu}$ and heat flux $F_{\rm{conv}}(T_{\mu})$ of a magma ocean below a stagnant lid. As the potential temperature decreases due to the cooling of the magma ocean, the viscosity increases, transitioning from values on the order of $10^{-1}$~Pa$\cdot$s to $10^{21}$~Pa$\cdot$s. With a potential temperature (T$_p$) of 1977~K, the temperature of the magma ocean crosses the liquidus, initiating crystallization at a pressure of $p = 0$~GPa. From this point, the function $f(\phi)$ drives the viscosity from $\mu_l$ to $\mu_s$ during the increase of the fraction of solids in the magma~\citep[see][]{costaetal09}. $T_{\mu}$ depends on the derivative of viscosity that is very sensitive to slope variations along the viscosity curve. Consequently, both $T_{\mu}$ and $F_{\rm{conv}}(T_{\mu})$ experience abrupt changes upon crossing the liquidus and solidus temperatures. In the calculations addressed in this work, the stagnant lid remains quite thin ($< 1$~km), so the thermal boundary between the lid and the magma ocean remains close to the surface, with $p \approx 0$~GPa.

\end{document}